\documentclass[usenatbib]{mn2e}
\usepackage{graphicx}
\usepackage{url}
\usepackage{amsmath}
\usepackage{array} % needed for H column type, see below
\usepackage{threeparttable}
\usepackage{hyperref}
\usepackage{caption3}
\usepackage[all]{hypcap}        % needed to help hyperlinks direct correctly

\usepackage{xcolor} 
\definecolor{darkblue}{rgb}{0,0,0.5}
\hypersetup{
  colorlinks   = true, %Colours links instead of ugly boxes
  urlcolor     = blue, %Colour for external hyperlinks
  linkcolor    = blue, %Colour of internal links
  citecolor    = darkblue   %Colour of citations
}

\newcolumntype{H}{@{}>{\lrbox0}l<{\endlrbox}}
\newcommand{\Ha}{H$\alpha$}
\newcommand{\UK}{$U$-to-$K$}
\newcommand{\pBC}{\emph{p}BC}
\newcommand{\p}{\emph{p}}
\newcommand{\dmf}{$\Delta m_{15,V}$}
\newcommand{\simlt}{{\small\raisebox{-0.6ex}{$\,\stackrel{\raisebox{-.2ex}{$\textstyle <$}}{\sim}\,$}}}
\newcommand{\simgt}{{\small\raisebox{-0.6ex}{$\,\stackrel{\raisebox{-.2ex}{$\textstyle >$}}{\sim}\,$}}}   

\newcommand{\leff}{$\lambda_{eff}$}

\begin{document}

\title[Bolometric corrections for CCSNe]{Bolometric Corrections for optical light curves of Core-collapse Supernovae}

\author[Lyman, Bersier \& James]
{\parbox{\textwidth}{J. D. Lyman$^1$\thanks{E-mail:
jdl@astro.livjm.ac.uk }, 
D. Bersier$^1$
and
P. A. James$^1$}\vspace{0.4cm}\\
$^1$Astrophysics Research Institute, Liverpool John Moores University, Liverpool, L3 5RF, UK\\
}

\date{Accepted . Received ; in original form }

\pagerange{\pageref{firstpage}--\pageref{lastpage}} \pubyear{2013}

\maketitle

\label{firstpage}

\begin{abstract}
Through the creation of spectral energy distributions for well-observed literature core-collapse supernovae (CCSNe), we present corrections to transform optical light curves to full bolometric light curves. These corrections take the form of parabolic fits to optical colours, with $B-I$ and $g-r$ presented as exemplary fits, while parameters for fits to other colours are also given. We find evolution of the corrections with colour to be extremely homogeneous across CCSNe of all types in the majority of cases, and also present fits for stripped-envelope and type II SNe separately. A separate fit, appropriate for SNe exhibiting strong emission due to cooling after shock breakout, is presented. Such a method will homogenise the creation of bolometric light curves for CCSNe where observations cannot constrain the emitted flux -- of particular importance for current and future SN surveys where typical follow-up of the vast majority of events will occur only in the optical window, resulting in consistent comparisons of modelling involving bolometric light curves. Test cases for SNe 1987A and 2009jf using the method presented here are shown, and the bolometric light curves are recovered with excellent accuracy in each case.
\end{abstract}


\section{Introduction}

Core-collapse supernovae (CCSNe) are spectacular highlights throughout the transient universe. As the extremely luminous end-point of the evolution of massive stars \citep[$\geq$~8~M$_{\odot}$;][]{smartt09}, CCSNe are valuable tools to many areas of astrophysical research. Their ability to probe the environments they inhabit and our understanding of the diverse range of explosions that occur, however, is limited by our knowledge of the progenitor system for each event, and the surrounding medium. 

If the event occurs in a local galaxy, searches in high resolution archival imaging can allow direct observations of the progenitor system to be made; such analysis has proved successful in a growing number of cases \cite[e.g.][see \citealt{smartt09} for a review]{vandyk03,smartt04,li07,galyam09,vandyk12a,maund11}. This technique is clearly reliant on the proximity of the SN, to be able to resolve individual progenitor systems from star clusters, and also the existence of archival high-resolution imaging to a sufficient depth for direct detection or stringent upper limits to be made (preferably in several bands). For the vast majority of discovered SNe, direct studies cannot be performed; post-explosion observations and modelling of the luminous event must be used to infer the properties of the progenitor star and the explosion. Dedicated SN searches are finding SNe of all types with such regularity that in-depth observational follow up, required for accurate modelling, is not feasible for most SNe discovered. 

Modelling of CCSNe, from simple analytical descriptions of the light curve evolution \citep{arnett82,valenti08} to spectral synthesis codes \citep[e.g.][]{mazzali93} and hydrodynamical modelling \citep[e.g.][]{nakamura01,utrobin07,tanaka09} can provide good estimates of the physical parameters of the explosion; typically the mass of nickel synthesised and the mass and kinetic energy of the ejecta. Such modelling typically requires at the very least one spectroscopic observation near peak (for analytical methods), with good spectroscopic coverage into the nebular phase desired to obtain the most accurate results from spectral/hydrodynamical modelling. However, to obtain accurate explosion parameters, models must typically be scaled to a bolometric light curve, this is particularly true for hydrodynamical modelling. 

A NUV-NIR light curve contains the vast majority of the light from a SNe, but obtaining well-sampled data over this wavelength range is expensive, especially for significant samples of objects. Data are often limited to much shorter wavelength ranges, meaning estimates of bolometric magnitudes can be vastly underestimating the emitted flux. \emph{(U)BVRI} integrated light curves have been used as a proxy for a bolometric light curve, although a comparable amount of flux is emitted in the NIR alone. Either no attempt is made to correct for flux outside this regime \citep[e.g.][]{young10,sahu11}, since no reliable methods exist, or a zero order assumption, that the fraction of flux outside the observed window is constant with time, is made \citep{elmhamdi11}, despite this demonstrably not being the case \citep[e.g.][Section\nobreakspace \ref {sect:fluxcont} of this paper]{valenti08, modjaz09}. An improved method is to find a similar object and assume the same proportional flux to be emitted outside the observed window \citep[][Schulze et al. in prep]{valenti08,mazzali13}. Bolometric light curves are therefore created using a variety of methods and this introduces uncertainties on how to compare the results of modelling consistently across events.

\citet{bersten09} have investigated bolometric corrections (BC) to three well-observed type II-P SNe (SNe~II-P) and two sets of atmosphere models. The progenitor stars of SNe~II-P are expected to be at the lower end of the mass range for CCSNe \citep{smartt09} and to have kept their outer layers throughout their evolution. These hydrogen-rich envelopes make their evolution well approximated by spherical explosions whose evolution is blackbody-like and continuum-dominated (until the end of the plateau phase). \citet{bersten09} indeed find very tight correlations between the  bolometric correction and optical colour of the SNe/models in their sample, providing a parameterised way of obtaining bolometric magnitudes from \emph{BVI} photometry. Including another well observed SN~II-P, \citet{maguire10} looked at bolometric corrections versus time, and found relatively similar evolution of the BC to $R$-band magnitudes between the four SNe, although the $V$-band BC appears rather more diverse.

\citet{pritchard13}, utilising \emph{Swift} data, have considered all CCSNe types to produce bolometric and ultraviolet corrections (UVC). The \emph{Swift} dataset is uniquely able to constrain the behaviour of SNe in the $\sim$1800--3000\ \AA{} wavelength regime. By correlating directly with optical colours and UV integrated fluxes, both taken with the UVOT instrument, they find a linear behaviour of the UVC, which appears to have no strong dependence on CCSN type. These correlations, however, are subject to substantial spread, which highlights the diversity of UV evolution in CCSNe. An attempt was also made to create a BC for CCSNe, however since the reddest filter available on UVOT is $V$, the BC is reliant upon modelling and the blackbody (BB) approximation for wavelength regimes that contain the majority of the bolometric flux at all but the very earliest epochs. As the authors noted, ground-based observations will provide a more robust estimate for the contribution of these longer wavelengths to CCSN bolometric light curves, particularly when including near infrared (NIR) observations. 

Clearly a consistent manner in which to obtain an approximation for the bolometric output, particularly for stripped-envelope CCSNe (SE~SNe; i.e.\ types Ib, Ic and IIb), is lacking. Such a method would allow results from modelling to be compared more consistently, as well as providing further tests for current and future models and simulations of SNe.

In this paper we utilise literature data for well observed CCSNe to investigate flux contributions of different wavelength regimes, and to construct BCs based on optical colours. Although these literature SNe are predominantly observed in the Johnson-Cousins systems, we also present fits in Sloan optical bands given their prevalence in current and future SN surveys. In Section\nobreakspace \ref {sect:data} we present the data and the SN sample, Section\nobreakspace \ref {sect:method} describes the steps involved in creating spectral energy distributions (SEDs) for the SN sample. Results and fit equations are presented in Section\nobreakspace \ref {sect:results} and discussed in Section\nobreakspace \ref {sect:discuss}.

\section{Data}
\label{sect:data}
\subsection{Photometry}
\label{sect:photometry}

SEDs, from which to calculate integrated light curves, would ideally be constructed from spectra. However the expense of such spectral coverage, and consequent dearth of available observations, means that the SEDs analysed here have been constructed exclusively from broad-band photometric data. The phase ranges covered by this analysis (typically $<$~70 days past peak for SE~SNe, and the duration of the plateau for type II SNe; SNe~II) are regions where continuum emission dominates the brightness of a SN, with line emission only dominating in the later, nebular phases. As such, photometric and spectroscopic integrated luminosities will typically agree well in the phase ranges explored here.

All photometric data used here are taken from the literature where a CCSN has photometric coverage over the \UK{} wavelength range. All types of CCSNe are included except those exhibiting strong interaction with their surrounding medium (typically with an `n' designation in their type). Strong CSM interaction introduces a range of photometric and spectroscopic evolution, as well as the possibility of early dust formation \citep[e.g. SN2006jc;][]{smith08,nozawa08}, making SED evolution between these events diverse.\footnote{Furthermore, recent evidence suggest some fraction of SN~IIn could be type Ia, thermonuclear, explosions that are expanding into a dense hydrogen-rich medium \citep{silverman13}.} See \citet{moriya13} for an analytical treatment of SNe~IIn bolometric light curves.

\subsection{SN sample}
\label{sect:sn_sample}
Naturally, a sample of well-observed SNe taken from the literature will be extremely heterogeneous since it is often the events that display unusual or peculiar characteristics (and/or are very nearby) that find the most attention. As such this sample is by no means a representative sample of discovered CCSNe, or CCSNe as a whole. This makes it more difficult to break down the sample by type as some are unique events. The unusual characteristics across this sample are evident from the uncertain and peculiar flags on their initial IAU typing.\footnote{\url{http://www.cbat.eps.harvard.edu/lists/Supernovae.html}} In Table\nobreakspace \ref {tab:sn_sample} we present SN type, as taken from more detailed literature studies of the objects, host galaxy name and redshift from the NASA Extragalactic Database (NED)\footnote{\url{http://ned.ipac.caltech.edu/}}, $E(B-V)$ values for Galactic and total reddening, the filters used to construct the SED, an epoch range over which the full filter set can be reliably used in constructing the SED (see Section\nobreakspace \ref {sect:interps}), and a \dmf{} value (see Section\nobreakspace \ref {sect:fluxcont}) for each SN in the sample. 

The sample includes a GRB-SN (SN1998bw), XRF-SNe (SN2006aj, SN2008D) and the unusual SN2005bf that displayed two peaks and a transition from type Ic to Ib, discussed variously as a magnetar \citep{maeda07} and an asymmetric Wolf-Rayet explosion \citep[e.g.][]{folatelli06}. See the references in Table\nobreakspace \ref {tab:sn_sample} for a detailed discussion of individual events and further unusual characteristics. 

Given the limited sample and the previously mentioned eclectic and peculiar nature of many of them, we limit our sub-typing to SE~SNe (i.e.\ those of type Ib, Ic and IIb) and SNe~II (i.e.\ those of any type II except IIb). Practically, we consider SN1987A, SN1999em, SN2003hn, SN2004et, SN2005cs and SN2012A as the SNe~II sample ($N=6$), with all others being SE~SNe ($N=15$).

\begin{table*}
\begin{threeparttable}
   \caption{Data for SNe in the sample.}
  \begin{tabular}{lllccccHccc}
  \hline
SN name  & Type &  Host         & Redshift & $E(B-V)_{MW}$ & $E(B-V)_{tot}$ &Filter coverage  & Photometric coverage & Full SED coverage\tnote{a} & \dmf{}\tnote{b}& Refs.\\
         &      &               &          &  (mag)        &    (mag)       &                 & ($\Delta t_{peak,V}$)&  ($\Delta t_{peak,V}$)  &  (mag)            &       \\
\hline
1987A    & II-pec&  LMC          &  0.0009  & 0.08   & 0.17   & \emph{UBVRIJHK} &  1--134\tnote{c}  & 2--134\tnote{c}       & ---   &  1--4    \\
1993J    & IIb   &  M81          & -0.0001  & 0.081  & 0.194  & \emph{UBVRIJHK}             &  $-$18--27        & $-$18--$-$10, 14--27  & 0.935 &  5--7    \\
1998bw   & Ic-BL &  ESO~184-G82  &  0.0087  & 0.065  & 0.065  & \emph{UBVRIJHK} &  $-$8--70         & 6, 31, 49             & 0.816 &  8,9     \\
1999dn   & Ib    &  NGC~7714     &  0.0093  & 0.052  & 0.10   & \emph{UBVRIJHK}             &  $-$1--123        & 24, 38, 123           & 0.500 &  10      \\
1999em   & II-P  &  NGC~1637     &  0.0024  & 0.043  & 0.10   & \emph{UBVRIJHK}             &  9--124\tnote{c}  & 11--117\tnote{c}      & ---   &  11,12    \\
2002ap   & Ic    &  M74          &  0.0022  & 0.072  & 0.09   & \emph{UBVRIJHK}             &  $-$8--25         & $-$8--25              & 0.881 &  13--20   \\
%2003bg   & IIb   &  MCG~-05-10-15&  0.0046  & 0.02   & 0.02   & \emph{BVRIJHK}              &  $\sim-$18--37    & $-$8--20              & 0.518 &  21     \\
2003hn   & II-P  &  NGC~1448     &  0.0039  & 0.014  & 0.187  & \emph{UBVRIYJHK}            &  20--170          & 20--140               & ---   &  12     \\
2004aw   & Ic    &  NGC~3997     &  0.0159  & 0.021  & 0.37   & \emph{UBVRIJHK}             &  $-$8-45          & 4--27                 & 0.558 &  21     \\
2004et   & II-P  &  NGC~6946     &  0.0001  & 0.314  & 0.41   & \emph{UBVRIJHK}             &  8--112\tnote{c}  & 8--112\tnote{c}       & ---   &  22,23    \\
2005bf   & Ib/c  &  MCG~+00-27-5 &  0.0189  & 0.045  & 0.045  & \emph{UBVriJHK}             &  $-$27--28\tnote{d}& $-$17--20\tnote{d}   & 0.462 &  24      \\
2005cs   & II-P  &  M51          &  0.0015  & 0.035  & 0.050  & \emph{UBVRIJHK}             &  3--93\tnote{c}   & 3--80\tnote{c}        & ---   &  25     \\
%2005ek   & Ic    &  UGC~2526     &  0.0165  & 0.210  & 0.03   & \emph{BVRIJHK}              &  0--18            & 1--16                 & ???   &  27,28   \\
2006aj   & Ib/c  &  Anon.        &  0.0335  & 0.142  & 0.142  & \emph{UBVRIJHK}             &  $-$9--14         & $-$7--6               & 1.076 &  26,27    \\
%2006jc   & Ib/c?&  UGC~4904     &  0.0056  & 0.02   & 0.05   & \emph{UBVRIJHK}             &  16--92\tnote{c}  & 50--92\tnote{c}       & ---   &  29--32   \\
2007Y    & Ib    &  NGC~1187     &  0.0046  & 0.022  & 0.112  & \emph{uBgVriYJHK}           &  $-$16--29        & $-$13--29             & 1.049 &  28      \\
2007gr   & Ic    &  NGC~1058     &  0.0017  & 0.062  & 0.092  & \emph{UBVRIJHK}             &  $-$8--141        & $-$3--141             & 0.861 &  29     \\  
2007uy   & Ib    &  NGC~2770     &  0.0065  & 0.022  & 0.63   & \emph{UBVRIJHK}             &  XXXXXXXX         & $-$4--5,33--35        & 0.815 &  30     \\
2008D    & Ib    &  NGC~2770     &  0.0065  & 0.023  & 0.6    & \emph{UBVRIJHK}             &  $-$17--18        & $-$16--18             & 0.697 &  31     \\  
2008ax   & IIb   &  NGC~4490     &  0.0019  & 0.022  & 0.4    & \emph{uBVrRIJHK}            &  $-$18--25        & $-$10--25             & 0.909 &  32,33    \\  
2009jf   & Ib    &  NGC~7479     &  0.0079  & 0.112  & 0.117  & \emph{UBVRIJHK}             &  $-$20--54        & $-$17--54             & 0.592 &  34,35    \\  
2011bm   & Ic    &  IC~3918      &  0.0015  & 0.032  & 0.064  & \emph{UBVRIJHK}             &  $-$19--78        & 8--56                 & 0.251 &  36     \\  
2011dh   & IIb   &  M51          &  0.0015  & 0.031  & 0.07   & \emph{UBVRIJHK}             &  $-$18--77        & $-$18--70             & 0.968 &  37     \\
2012A    & II-P  &  NGC~3239     &  0.0025  & 0.028  & 0.037  & \emph{UBVRIJHK}             &  3--140\tnote{c}  & 9--90\tnote{c}        & ---   &  38      \\
%PTF12gzk & Ic    &  SDSS~J221241.53+003042.7 & 0.0137 & 0.054 & 0.14 & \emph{BVRIJHK}       &  $-$12--4         & $-$5--1               & ???   &  41  \\
\hline
\end{tabular}
\label{tab:sn_sample}
\begin{tablenotes}
 \item [a]{The phase(s) over which there exists a full complement of filter observations (or well constrained interpolations) from which to construct an SED in days relative to the $V$-band peak}
 \item [b]{Difference in magnitudes of the $V$-band light curve} at peak and 15 days later
 \item [c]{Phase is quoted with respect to estimated explosion date}
 \item [d]{The second $V$-band peak is used as $t_{peak}$, SN2005bf is the famous `double-humped' SN.}
\end{tablenotes}
\vspace{0.3cm}
References:
(1) \citet{menzies87}; (2) \citet{catchpole87}; (3) \citet{gochermann89}; (4) \citet{walker90};
(5) \citet{richmond94}; (6) \citet[][and IAU circulars within]{matthews02}; (7) \citet{matheson00}; 
(8) \citet{clocchiatti11}; (9) \citet{patat01};
(10) \citet{benetti11};
(11) \citet{elmhamdi03}; (12) \citet{krisciunas09};
(13) \citet{mattila02}; (14) \citet{hasubick02}; (15) \citet{riffeser02};(16) \citet{motohara02}; (17) \citet{galyam02}; (18) \citet{takada02}; (19) \citet{yoshii03}; (20) \citet{foley03};
%(21) \citet{hamuy09};
% krisciunas09 again
(21) \citet{taubenberger06};
(22) \citet{zwitter04}; (23) \citet{maguire10};
(24) \citet{tominaga05};
(25) \citet{pastorello09};
%(27) \citet{modjaz07}; (28) \citet{drout13};
(26) \citet{mirabal06}; (27) \citet{kocevski07};
%(29) \citet{pastorello07}; (30) \citet{dicarlo08}; (31) \citet{mattila08a}; (32) \citet{anupama09};
(28) \citet{stritzinger09};
(29) \citet{hunter09};
(30) \citet{roy13};
(31) \citet{modjaz09};
(32) \citet{taubenberger11}; (33) \citet{pastorello08};
(34) \citet{valenti11}; (35) \citet{sahu11};
(36) \citet{valenti12};
(37) \citet{ergon13};
(38) \citet{tomasella13}.
%(41) \citet{benami12}.

\end{threeparttable}
\end{table*}

\section{Method}
\label{sect:method}

Flux evolution and BCs are found through integrations of various wavelength regimes of SEDs for our SN sample. A description of how these SEDs are constructed from the photometric data, and the treatment of unconstrained wavelengths follows.

\subsection{Interpolations of light curves}
\label{sect:interps}

Photometric data will not have equal sampling across all filters. For example optical data may be taken on a different telescope to the NIR, or poor weather prevent the observations in one or more bands on a given night. Since we are interested in obtaining a full SED over the \UK{} filter range, we must rely on interpolations in order to provide good estimates for these missing data.

Such interpolations were fitted to each filter light curve as a whole and chosen as the best estimate of the missing evolution of the light curve. Typically interpolation functions were either linear, spline or a composite fit \citep[consisting of an exponential rise, a Gaussian peak, and magnitude-linear decay; see][]{vacca96}. The choice of function was linked to the sampling; where the light curve had densely sampled evolution ($\sim$ daily), linear interpolation was sufficient, whereas splines and the composite model were used when the light curve had substantial gaps (\simgt{} several days) where the light curve was not constrained.

Interpolated values were used to fill in missing values from literature photometry such that at every epoch of observation a full complement of magnitudes in each filter of the \UK{} range existed from a mixture of observed and interpolated data points. Epochs over which the interpolations were valid were noted and interpolated values were only trusted within a few days of observations; for regions of simple behaviour, where we could be confident the interpolation accurately represented the missing part of the light curve (e.g.\ epochs on the plateau for SNe~II), this limit was increased. Any epoch where the evolution of the SN light curve in one or more filters was not well constrained was rejected from further analysis. Extrapolations were typically not relied upon, although some cases warranted extrapolated magnitudes to be used in one or two filters -- these were only used \simlt{}2 days beyond the data, and where the function was well-behaved. See Table\nobreakspace \ref {tab:sn_sample} for ranges where full \UK{} fluxes could be used in SED construction for each SN.

\subsection{SED construction}
\label{sect:sed}

SED construction is performed using a different method for three different wavelength regimes: The optical-NIR (3659--21900\ \AA{}; the wavelength range covered by the \UK{} photometry), the BB tail ($>$21900\ \AA{}) and the UV ($<$3569\ \AA{}). A discussion of the construction of the SED in each regime follows.

\subsubsection{The optical-NIR regime}
\label{sect:sed_opt}
In this wavelength range we are constrained by photometric observations from the interpolated light curves, which form tie points of the SED. Prior to SED construction, the photometry is corrected for extinction assuming a \citet{fitzpatrick99}\footnote{The choice of extinction law has minimal impact on the final results.} $R_V=3.1$ Galactic extinction curve for both Milky Way and host galaxy extinction. $E(B-V)_{tot}$ ($= E(B-V)_{MW} + E(B-V)_{host}$) values are given in Table\nobreakspace \ref {tab:sn_sample}. 

Extinction-corrected magnitudes are then converted to fluxes ($F_{\lambda}$). An optical-NIR SED is created for every epoch of observation using the $F_{\lambda}$ and effective wavelength (\leff{}) values of each filter. Filter zeropoints, to convert to $F_{\lambda}$, and \leff{} values are taken from \citet{fukugita96}, \citet{bessell98} and \citet{hewett06}. Note that $K$-corrections were neglected in this analysis due to the very low redshift of the sample (see Table\nobreakspace \ref {tab:sn_sample}). $K$-corrections were investigated using available spectra of the SN sample at similar epochs to SED construction in WISeREP \citep{yaron12}, with over 90 per cent of measurements across all filters having $|K| < 0.03$~mag.

\subsubsection{The BB tail}
\label{sect:sed_ir}

Although longer wavelengths than $K$-band are not expected to contribute significantly to the bolometric flux, a treatment of these wavelengths in the SEDs must be made to avoid systematically underestimating the bolometric flux. During the photospheric epochs mainly investigated here, we assume the flux evolution of the long wavelength regime to be well described by a BB tail -- see Section\nobreakspace \ref {sect:uvir_treatment} for a discussion of this approximation.

A BB was fit using the $R$ (or $r$), $I$ (or $i$), $J$, $H$ and $K$-band fluxes ($R$-to-$K$), since optical fluxes, particularly for SE~SNe, fall below the expectation from a BB once strong line development of Fe-group elements begins \citep[see][and references therein]{filippenko97}. Epochs where bluer bands are expected to be well characterised by a BB fit (\simlt{}20 days for SE~SNe and prior to the end of the plateau for SNe~II), were also separately fitted, here including $B$- and $V$-bands, to ascertain the difference to the $R$-to-$K$ fits. Including these extra bands had very little impact on the fits and resulting integrated luminosities. As such we favour using the $R$-to-$K$ bands for our BB fits, since this is appropriate for each SNe at all epochs investigated here and we reduce the danger of erroneously fitting to wavelengths that are not described by a BB.

{\sc curve\_fit} in the {\sc Scipy}\footnote{\url{http://www.scipy.org/}} package was used on each pair of parameters in an initial grid of reasonable SN temperatures and radii to find the global $\chi^2$ minimised BB function. The resulting function was appended to the optical-NIR SED at the red cut off of the $K$-band filter (defined as 10 per cent transmission limit, 24400\ \AA{}) and extended to infinity. The $K$-band and beginning of the BB tail were linearly joined in the SED.

\subsubsection{The UV}
\label{sect:sed_uv}
The UV represents a wavelength regime with complex and extremely heterogeneous evolution for CCSNe \citep{brown09}. Coupled with a dearth of observations, correcting for flux in the UV is uncertain. Early epochs in the evolution of a SN can be dominated by the cooling of shocked material which emerges after the short-lived shock breakout (SBO) emission. This cooling phase is observed as a declining bolometric light curve that is very blue in colour. After this the radioactively powered component of the light curve begins to dominate and the light curve then rises to the radioactive peak (in SE~SNe; for SNe~II-P the recombination-powered light curve will become dominant and the light curve will settle to the plateau phase). The time over which the cooling phase dominates is highly dependent on the nature of the progenitor star, primarily driven by its size. The extended progenitors of SNe~II, which have retained their massive envelopes, can display the signature of this cooling phase for many days, whereas in compact SE~SNe progenitors it is shorter and often \simlt{}1~day. Indeed for SE~SNe it has only been seen in a handful of cases (e.g. SNe 1993J, \citealt{richmond94}; 1999ex, \citealt{stritzinger02}; 2008D, \citealt{modjaz09}; 2011dh, \citealt{arcavi11}), generally thanks to extremely early detections.
The evolution in the UV regime also quickly falls below the expectations of a BB approximation, as mentioned in Section\nobreakspace \ref {sect:sed_ir}, and as such a BB fit to these wavelength ranges over most of the evolution of a SN would be inconsistent with one drawn from longer wavelengths. During the cooling phase however, a BB fit across all wavelengths is appropriate as the SN is dominated by the hot, continuum flux. 

Given the changing behaviour of the UV we utilise two treatments for the differing cases.
For epochs over the cooling phase, the UV flux is taken to be the integrated flux of a BB function, from zero\ \AA{} to the blue edge of the $U$-band \citep[following e.g.][]{bersten09}; the BB is fitted and joined to the SED in the same manner as Section\nobreakspace \ref {sect:sed_ir}. (Note that for these epochs we opted to include the $B$ and $V$ filters as further constraints for the BB.) Signatures of this cooling phase were taken to be early declines in the $U$- and $B$-bands in the light curves of SNe. All SNe~II and SNe 1993J, 2008D and 2011dh had epochs during the cooling phase which contained full \UK{} photometry, i.e.\ where we could contruct SEDs for them. The extent of the cooling phase was determined by observing a drop in the $U$-band flux in the SED, relative to that predicted by the BB fit.

To account for UV flux at later epochs, when the BB approximation is not appropriate, each SED was tied to zero flux at 2000\ \AA{} by linearly extrapolating from the $U$-band flux. This was found to be a good estimate of the UV flux when compared to UV observations, as discussed in Section\nobreakspace \ref {sect:uvir_treatment}.  

\section{Results}
\label{sect:results}

Using the constructed SEDs, investigations into the contributions of different wavelength regimes can be made over various epochs of SN evolution and across different types.

\subsection{Flux contributions with epoch}
\label{sect:fluxcont}

\begin{figure}
 \includegraphics[width=\linewidth]{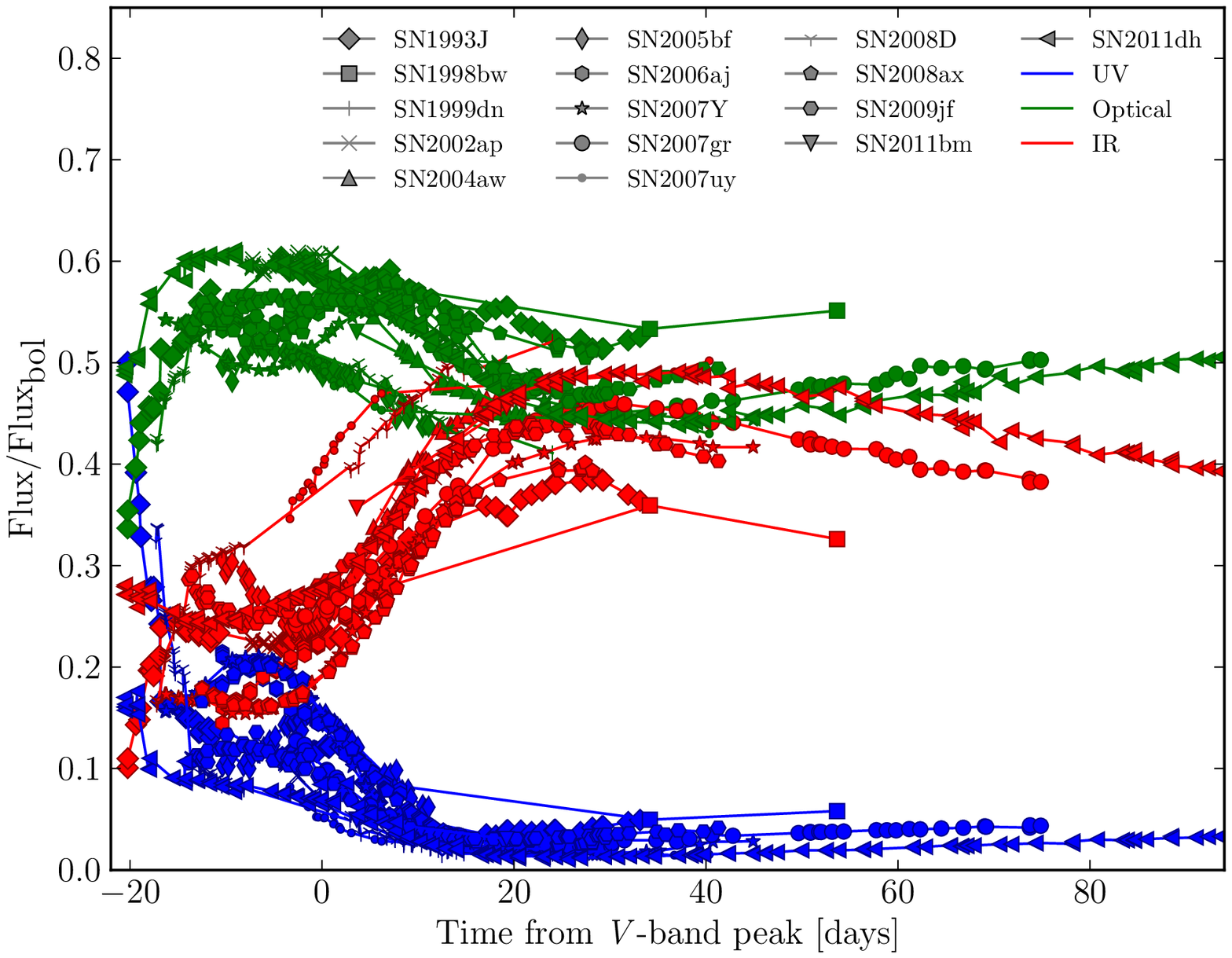}
  \includegraphics[width=\linewidth]{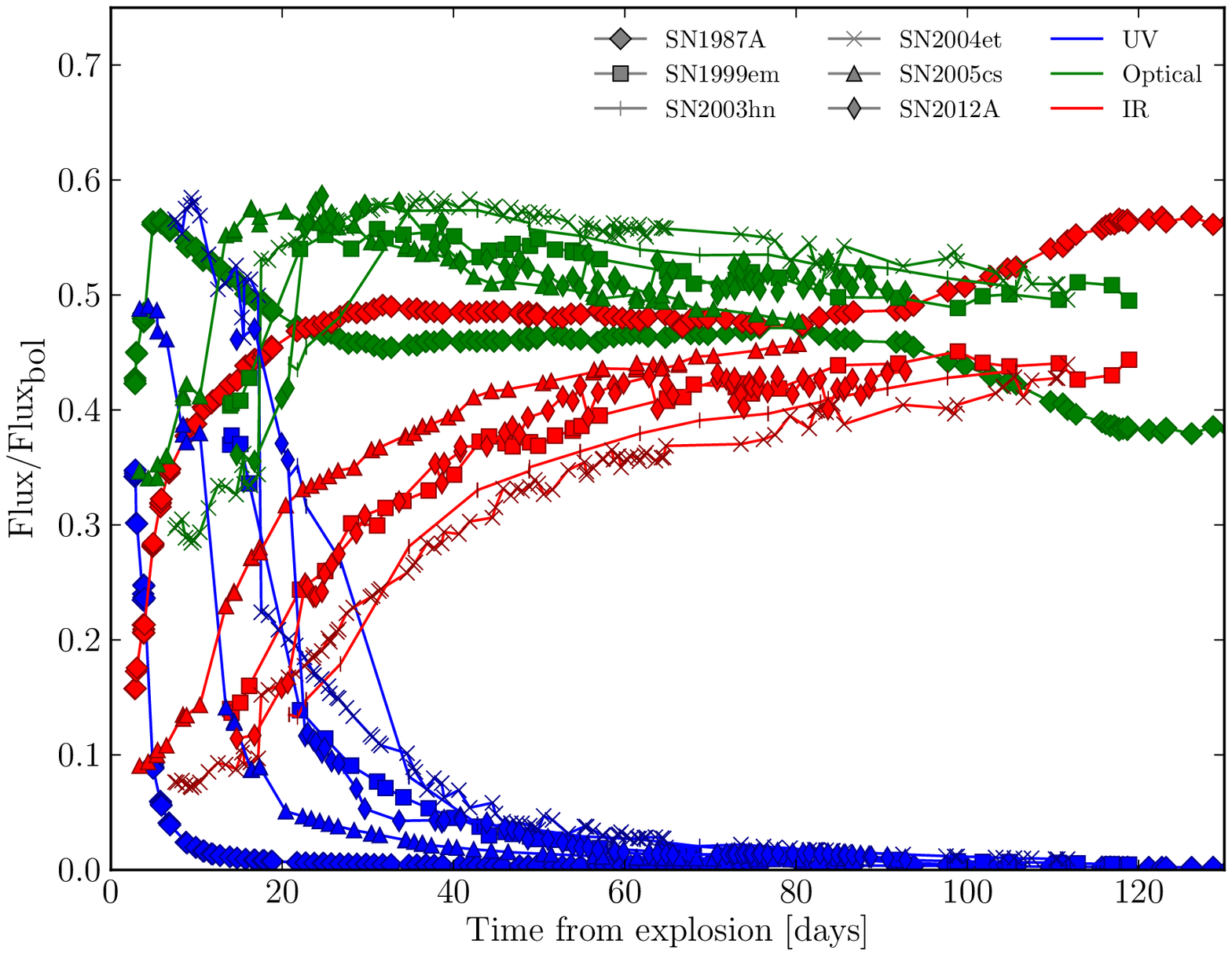}
 \caption{The evolution with time of the UV (blue), optical (green), and IR (red) fluxes as a fraction of the bolometric flux for SE~SNe (top) and SNe~II (bottom). Light curve evolution has been stretched for each SE~SNe to a common \dmf{} of 0.758 (see text).}
 \label{fig:fluxvstime}
\end{figure}

Initially, SEDs were integrated over three wavelength regimes: an optical regime (defined here as covering the \emph{gri} bands; 3924--8583\ \AA{}), the IR (including the BB tail; 9035\ \AA{}--infinity) and the UV (0--3924\ \AA{}). For the SE~SNe each SED was assigned an epoch relative to the peak of the $V$-band light curve, where the peak was found by fitting a polynomial to the data near maximum. A \dmf{} value for each SN was also found, following the method of \citet{phillips93}, using the $V$-band. \dmf{} values are presented in Table\nobreakspace \ref {tab:sn_sample}. Each SN was normalised to the evolution of the average \dmf{} (0.758) in order to correct for varying light curve evolution time-scales. This was done by applying a linear stretch factor, $S$ to the epoch for each SN, where $S = \Delta m_{15,V}/0.758$. For SNe~II each SED epoch was made relative to the time of explosion (see references in Table\nobreakspace \ref {tab:sn_sample}).

The behaviour of the flux contained in the UV, optical, and IR are plotted as a function of the total bolometric flux in Fig.\nobreakspace \ref {fig:fluxvstime}. Clearly the individual components, as fractions of the total emitted flux, evolve strongly with respect to time, as has been indicated previously for individual or small samples of events \citep[e.g.][]{valenti08,modjaz09,stritzinger09}. This behaviour appears to be qualitatively similar for all SE~SNe (after normalising to a common light curve evolution time-scale). IR fractions typically reach a minimum on or slightly before $V$~peak and then rise until $\sim 20/S$ days after peak, reaching a comparable fraction of the bolometric flux to that of the optical. The UV is weak at most epochs, falling from $\sim$10--20 per cent prior to peak to just a few per cent a week past peak for most SE~SNe. However, in the case of an observed SBO cooling phase, as is the case particularly for SN1993J and, to a lesser extent, SNe 2008D and 2011dh, the UV contributes a significant fraction of the bolometric flux for a short time after explosion. Due to the putative compact nature of the progenitors of these SNe however, the fraction of the light emitted in the UV falls rapidly.

For SNe~II we see largely coherent behaviour amongst the sample in the three regimes, albeit very different from that of SE~SNe. The IR rises almost monotonically from explosion until the end of the plateau, where it contributes $\sim$40 per cent to the bolometric flux. The optical remains roughly constant with time indicating the BC to optical filters should be roughly constant \citep[e.g.][]{maguire10}. The UV contributes a larger fraction than in SE~SNe and for a longer time, owing largely to the generally much more extended cooling phase that SNe~II exhibit, for example SN2003hn shows significant UV contributions to its bolometric flux ($\sim$30~per~cent) more than 20 days past explosion. SN1987A, however, is very unusual compared to the other events. Being UV deficient \citep{danziger87}, any significant contribution from the cooling phase rapidly falls, with the UV making up only a few percent of the bolometric flux within a week of explosion. The IR of SN1987A also increases much more rapidly than other SNe~II, maintaining a similar fraction as that of the optical from 20~days, and overtaking the optical as the dominant regime after $\sim$80~days.

\subsection{Optical colours and bolometric corrections}
\label{sect:optcols}

CCSNe evolve strongly in colour during the rise and fall of their brightness. Previous work looking at the optical colours of SE~SNe and SNe~II \citep[e.g.][]{drout11,maguire10} shows that the colour evolution  
changes strongly as a function of time.
The driving force of these large colour changes during the photospheric evolution of a SN is the change in temperature of the photosphere, with some smaller contribution from development of heavy element features in the spectra. 
It is  expected that the BC should be linked to the colour of the SN (a diagnostic for the temperature) at that epoch. 
Given the relative ease of obtaining colours for SNe as oppose to characterising entire flux regimes as is given in Section\nobreakspace \ref {sect:fluxcont}, it is prudent to quantify BCs as functions of the colours sensitive to the SN's temperature (i.e.\ those in the optical regime).

For filters used in the construction of the SEDs, obtaining the colour at each epoch is trivial. However, when one or both filters are not observed and thus do not form tie points of the flux in the SEDs, we must rely on interpolations. The linear SED interpolations used in order to integrate over wavelength were used to sample the SEDs at the \leff{} of the desired filter, and fluxes in $F_\lambda$ were then converted to apparent magnitudes. The continuum-dominated SEDs largely do not contain significant fluctuations on the scale of broadband filter widths between neighbouring broadband filters and one would not expect large deviations from a linear interpolation between neighbouring filters. In the interest of presenting results that will be useful for future surveys, corrections to Sloan magnitudes were investigated. An analysis of using these linear interpolations to derive Sloan magnitudes is made in Appendix\nobreakspace \ref {sect:extractsloan} by comparing to the expected magnitudes directly from contemporaneous spectra. We find that $g$ and $r$ magnitudes are very well estimated by the linear interpolation method, however there is some systematic offset in $i$.

Given the highly uncertain nature of the UV correction, two types of BC were investigated. These are a `true' BC including the UV, and what will be termed a \emph{pseudo-}BC (\pBC{}) which will neglect contributions from the UV (i.e.\ the BB integration to zero\ \AA{} or linear extrapolation to 2000\ \AA{}, see Section\nobreakspace \ref {sect:sed_uv}) and instead cut off at the blue edge of the $U$-band. This makes the \pBC{} independent of the treatment of the UV presented here and makes no attempt to account for these shorter wavelengths, useful in the case where UV observations exist, where indications of unusual UV behaviour are present, or where a complementary treatment of the UV exists that may be added to the \pBC{}.

The SEDs were integrated over each of the wavelength ranges to obtain (pseudo-)bolometric fluxes. These were then converted to luminosities, and finally to bolometric (or pseudo-bolometric) magnitudes using:
\begin{equation}
  M_{bol} = M_{\odot,bol}-2.5\log_{10}\left(\frac{L_{bol}}{L_{\odot,bol}}\right),
\label{eq:bolo}
\end{equation}
where $M_{bol}$ and $L_{bol}$ can be replaced by their pseudo-bolometric counterparts. A BC (or \pBC{}) to filter $x$ can then be defined as:
\begin{equation}
  \text{BC}_{x} = M_{bol} - M_{x},
\label{eq:bc}
\end{equation}
where $M_{x}$ is the absolute magnitude of SN in filter \emph{x} that has been corrected for extinction (see Section\nobreakspace \ref {sect:sed_opt}). This definition can also be expressed in observed magnitudes \mbox{($\text{BC}_{x} = m_{bol} - m_{x}$)} using the distance modulus for each SN host.\footnote{Although the BC is accounting for missing flux, its value can be positive in magnitudes, given the difference in the zeropoints for the filter magnitudes and $M_{bol}$.}

All colours and (\p)BCs in the $BVRI$ and $gri$ ranges were computed. For both the \pBC{} and BC, the tightest correlation for the Johnson-Cousins filters was $BC_{B}$ against $B-I$ colour. For the Sloan filters this was the $BC_{g}$ against $g-i$ colour, however, as detailed in Appendix\nobreakspace \ref {sect:extractsloan}, the $i$-band derived magnitudes are susceptible to a systematic offset, and as such we present $g-r$ as the representative fit.

We will limit our discussion here to mainly the BC to $B-I$, alongside plotting the BC to $g-r$ relation for a visual comparison; the parameters for all reasonable \pBC{} and BC fits, which may be useful in the case where good coverage is not available in either of these filter pairs, are presented in Section\nobreakspace \ref {sect:otherfits}. 

Furthermore, distinct behaviour was observed for those epochs during the cooling phase (see Section\nobreakspace \ref {sect:sed_uv}) and subsequent epochs, mainly due to the differing behaviour of the UV and subsequent differing treatment in our method. We thus present the two phases separately and offer distinct fits to each.

\subsection{The radiatively/recombination powered phase}
\label{sect:radphase}

\begin{figure}
 \centering
 \includegraphics[width=\linewidth]{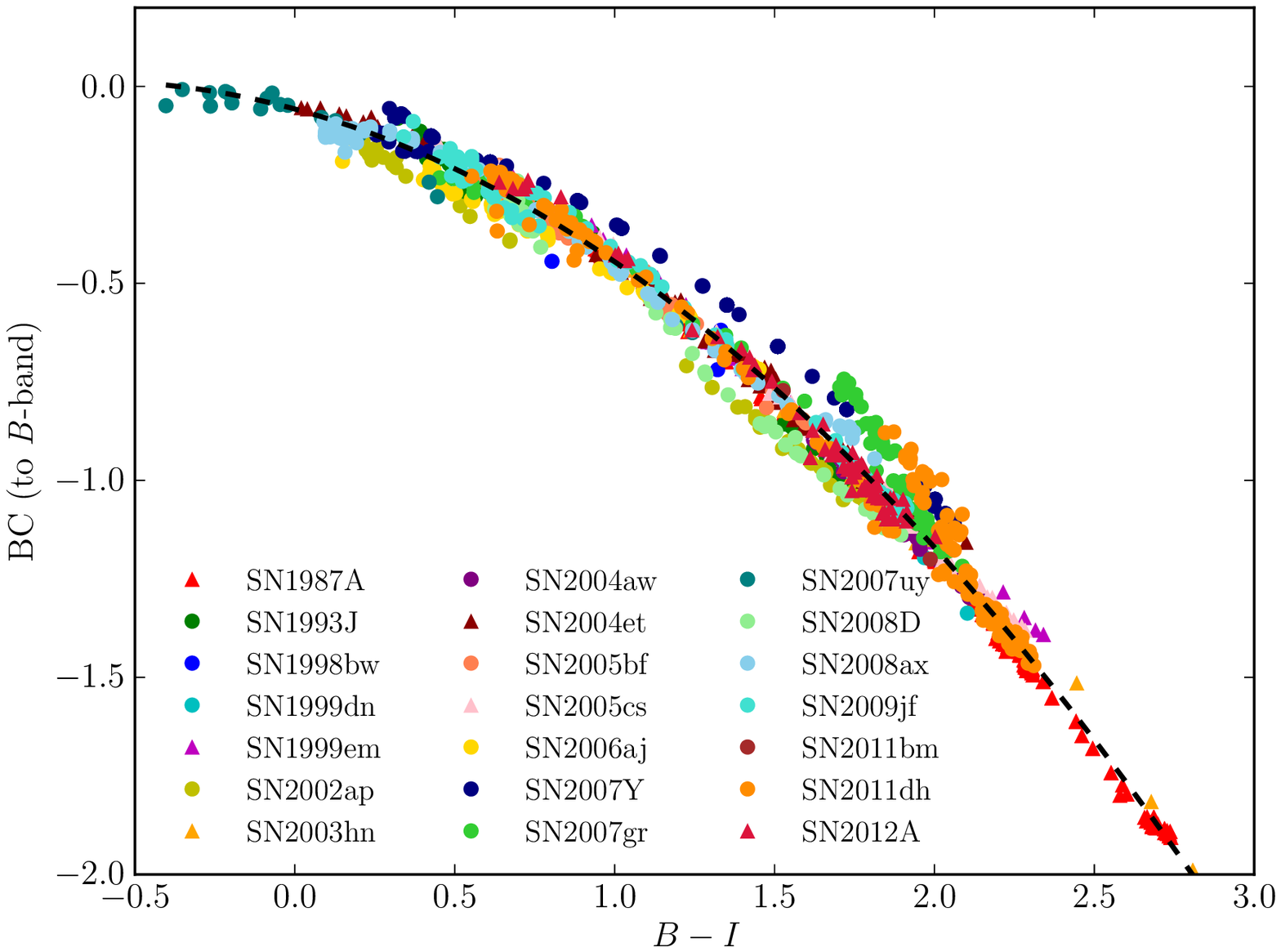}
 \includegraphics[width=\linewidth]{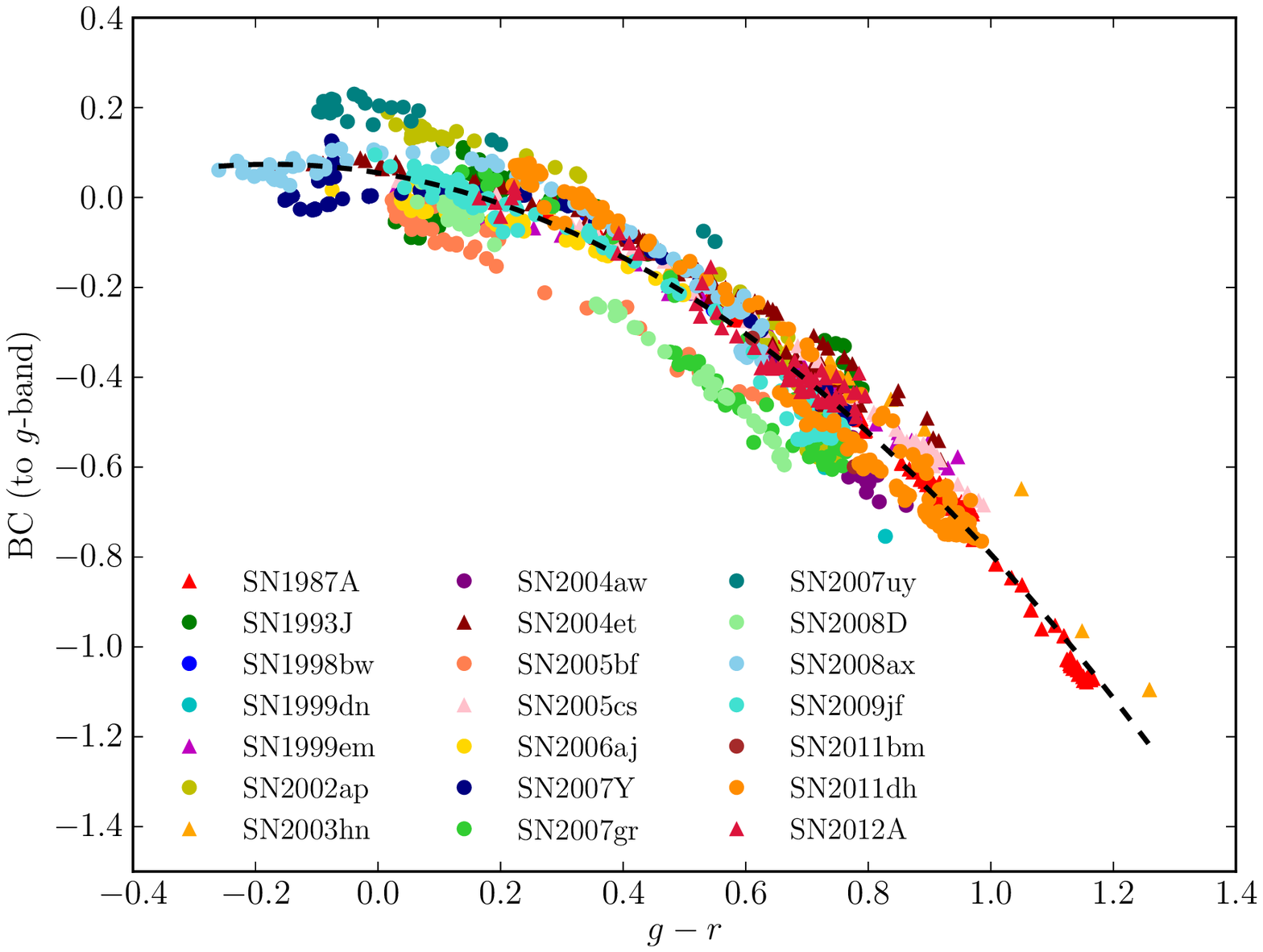}
\caption{BC for all SNe in the sample presented for a relation in the Johnson-Cousins (top) and Sloan (bottom) filters. Epochs shown \emph{do not} include those exhibiting signatures of strong cooling after SBO. SNe~II are denoted by triangles. A best-fitted second order polynomial is shown for each (see text).}
\label{fig:bc_all}
\end{figure}

\begin{figure}
 \centering
 \includegraphics[width=\linewidth]{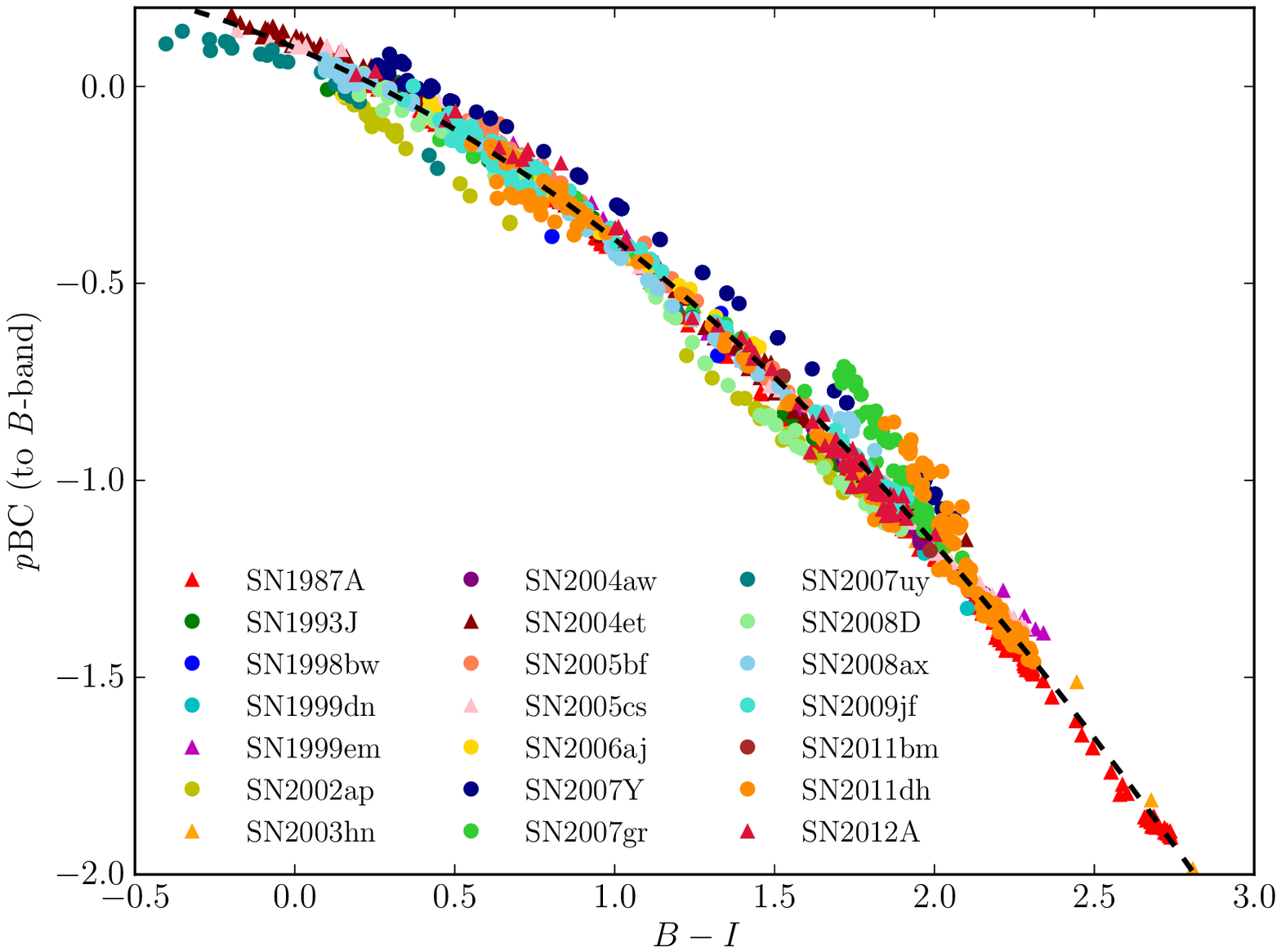}
 \includegraphics[width=\linewidth]{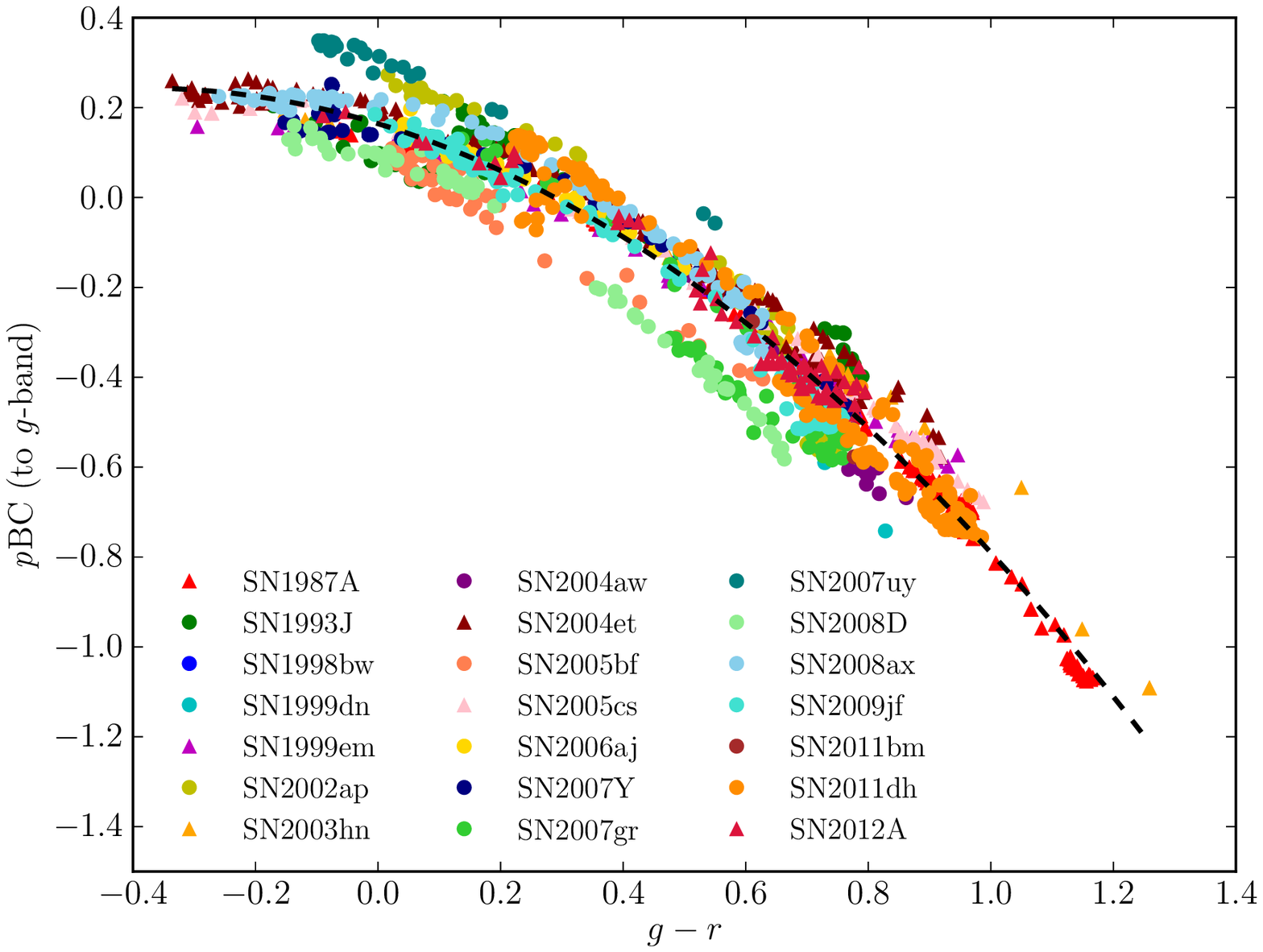}
\caption{\pBC{} for all SNe in the sample presented for a relation in the  Johnson-Cousins (top) and Sloan (bottom) filters. Epochs shown include those exhibiting signatures of strong cooling after SBO, since the UV is not accounted for in the \pBC{}. SNe~II are denoted by triangles. A best-fitted second order polynomial is shown for each (see text).}
\label{fig:pbc_all}
\end{figure}

Those epochs post cooling from SBO are analysed here. The $B-I$  and $g-r$ data for these epochs are plotted in Fig.\nobreakspace \ref {fig:bc_all} for the BC and Fig.\nobreakspace \ref {fig:pbc_all} for the \pBC{}.\footnote{All plots of BC against a given colour have equal plotted ranges. for ease of comparison.} As is evident, even across all SNe types, we find a tight correlation between the (\p{})BC, and the respective colour. Such a universal trend of behaviour allows us to construct fits to describe the bolometric evolution of CCSNe for each filter set. The BC has some parabolic evolution evident at blue and very red epochs, and as such a second order polynomial is fitted for the BC in each case.

Equations\nobreakspace \textup {(\ref {eq:bc_all_jc})} and\nobreakspace  \textup {(\ref {eq:bc_all_sl})} describe the BC fits to the entire sample, which allow a good estimate of a SN's bolometric magnitude to be made based on the colour in each equation.
\begin{equation}
  \text{BC}_{B} = -0.057 - 0.219 \times (B-I) - 0.169 \times (B-I)^{2} 
\label{eq:bc_all_jc}
\end{equation}
\begin{equation}
  \text{BC}_{g} = 0.055 - 0.219 \times (g-r) - 0.629 \times (g-r)^{2}
\label{eq:bc_all_sl}
\end{equation}

The BC in each case is a tight correlation, with deviations of just $\sim$0.1~mag from the best fitted function for even the most extreme objects in each case. The rms scatter and colour range for the $B-I$ fit are 0.053~mag and $-$0.4--2.8. For the $g-r$ fit the rms and colour range are  0.070~mag and $-$0.3--1.2.

\begin{figure}
 \centering
  \includegraphics[width=\linewidth]{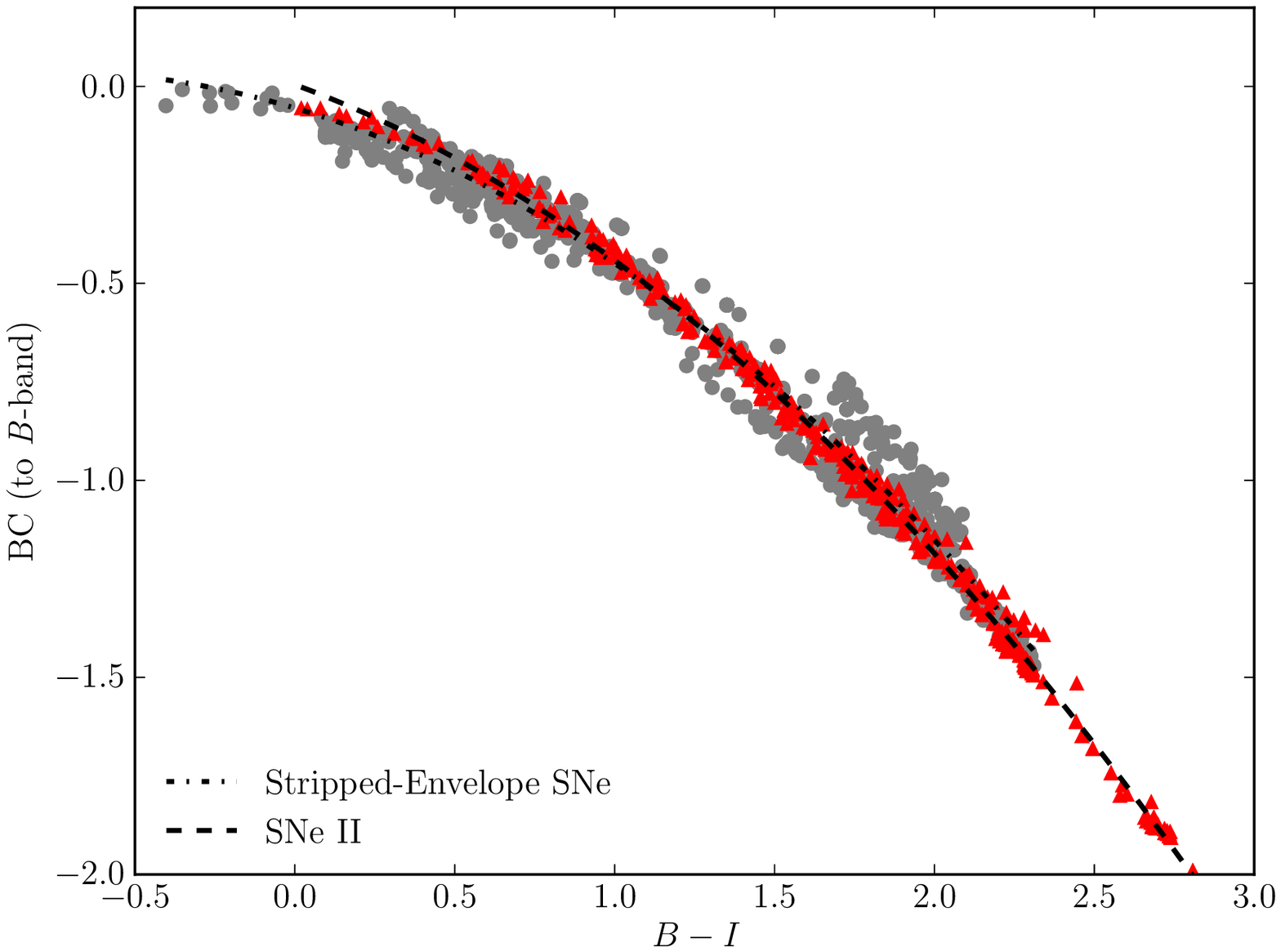}
 \includegraphics[width=\linewidth]{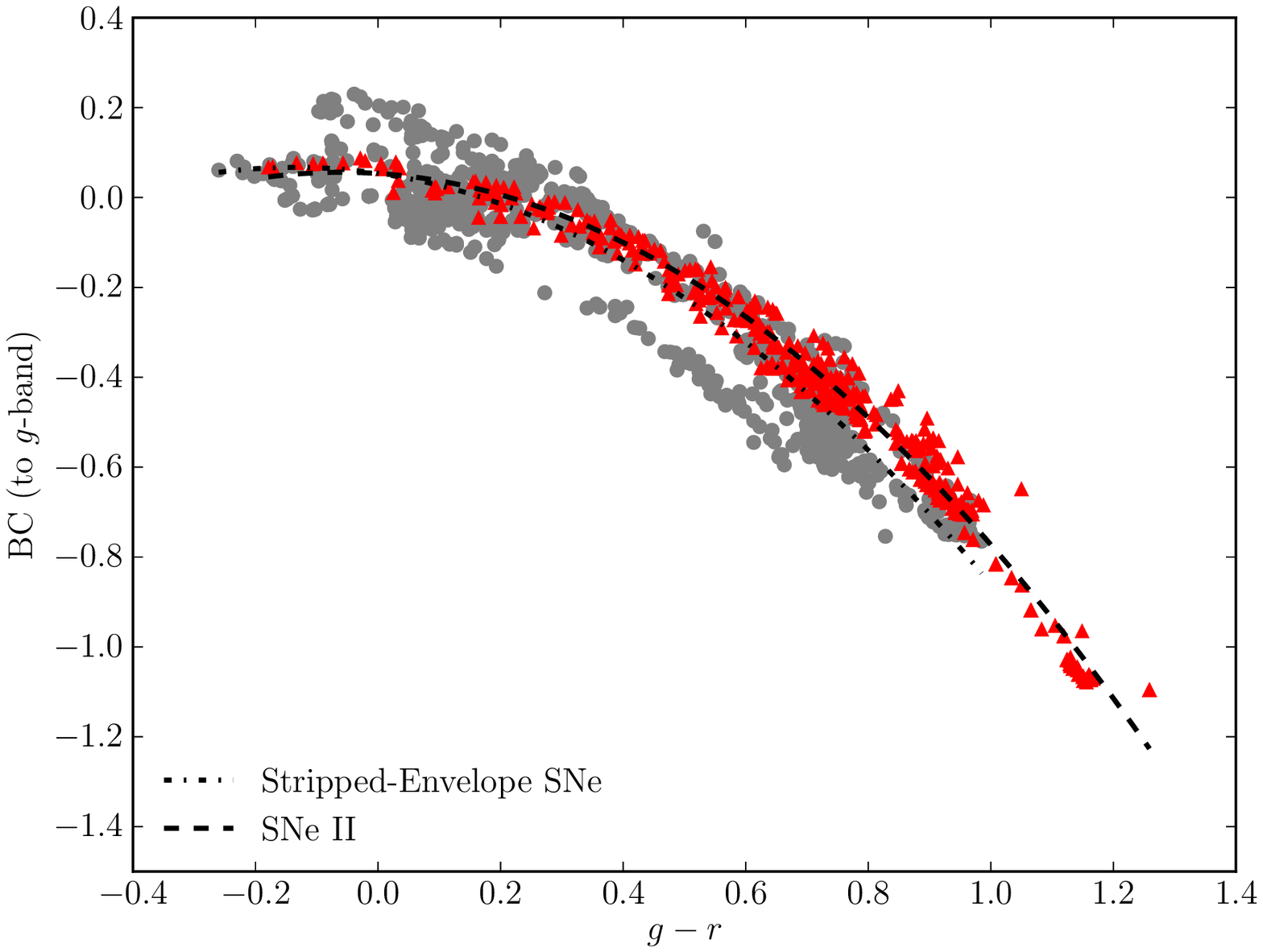}
 \caption{As for Fig.\nobreakspace \ref {fig:bc_all} but SNe are split between SNe~II (red triangles) and SE~SNe (Ib/c, IIb; grey circles) with a best-fitting second-order polynomial constructed for each SN type. Epochs shown \emph{do not} include those exhibiting signatures of strong cooling after SBO.}
 \label{fig:bc_twotype}
\end{figure}

Despite the generally universal behaviour of the SNe in the sample, there is certainly a difference in scatter and colour range between the two types, with SNe~II populating very red regions of the plots, and, although there is no indication for a strong divergence of the SE~SNe from the extended behaviour of the SNe~II, each SN type should only be trusted over the observed colour range. For these reasons, it is useful to define individual fits for SE~SNe and SNe~II separately. These are plotted in Fig.\nobreakspace \ref {fig:bc_twotype} for each filter set, colour-coded by type and with the individual fits to SE~SNe and SNe~II shown. As is clear from these fits, there is good agreement between the two samples over the range of colours for which the samples overlap.

The equations describing the SE~SN-sample fits shown in Fig.\nobreakspace \ref {fig:bc_twotype} are:
\begin{equation}
  \text{BC}_{B} = -0.055 - 0.240 \times (B-I) - 0.154 \times (B-I)^{2}
\label{eq:bc_se_jc}
\end{equation}
\begin{equation}
  \text{BC}_{g} = 0.054 - 0.195 \times (g-r) - 0.719 \times (g-r)^{2}
  \label{eq:bc_se_sl}
\end{equation}

\noindent
And for the SNe~II sample, the fits are:
\begin{equation}
\text{BC}_{B} = 0.004 - 0.297 \times (B-I) - 0.149 \times (B-I)^{2}
\label{eq:bc_ii_jc}
\end{equation}
\begin{equation}
  \text{BC}_{g} = 0.053 - 0.089 \times (g-r) - 0.736 \times (g-r)^{2}
\label{eq:bc_ii_sl}
\end{equation}

As might be expected given their more homogeneous evolution, SNe~II appear to evolve extremely similarly (including SN1987A, which displayed a very unusual light curve) until the end of the plateau, the time range over which this analysis is made. This confirms the coherent behaviour of SNe~II-P shown by \citet{bersten09} and indicates colour is a very good indicator of the BC for SNe~II. We present the bolometric light curve  of SN1987A constructed using the fits of \citet{bersten09} and those presented here in Appendix\nobreakspace \ref {sect:09jf}. We find a simple second-order polynomial sufficient to define the BC from our colours with a larger sample (up until the end of the plateau), which means the bolometric light curve of a SN~II can be robustly estimated from just two-filter observations with minimal scatter in the relation. An increase in sample size is obviously desired to improve and confirm this relation across the family of SNe~II. 

SE~SNe are an inherently diverse range of explosions given their various expected progenitor channels; notwithstanding this, we still see evolution remarkably well described by a second-order polynomial in each colour. Rather the opposite of investigating a ``typical'' SE~SNe sample, we here show many unique and unusual outbursts, which suggests that the spread observed here is plausibly close to the worse-case scenario of uncertainties on computing a bolometric magnitude for a given SE~SN that is constrained only in the optical. SN2007uy appears as somewhat of an outlier from the SE~SNe fit and this SN is discussed further in Section\nobreakspace \ref {sect:uvir_treatment}.

For the fits to $B-I$ the rms values are 0.061 and 0.026~mag for SE~SNe and SNe~II, respectively. The SE~SN (SN~II) fit is valid over the $B-I$ colour range $-$0.4--2.3 (0.0--2.8).

For the fits to $g-r$ the rms values are 0.076 and 0.036~mag for SE~SNe and SNe~II, respectively. The SE~SN (SN~II) fit is valid over the $g-r$ colour range $-$0.3--1.0 ($-$0.2--1.3)

\subsection{The cooling phase}
\label{sect:coolingphase}

\begin{figure}
 \centering
 \includegraphics[width=\linewidth]{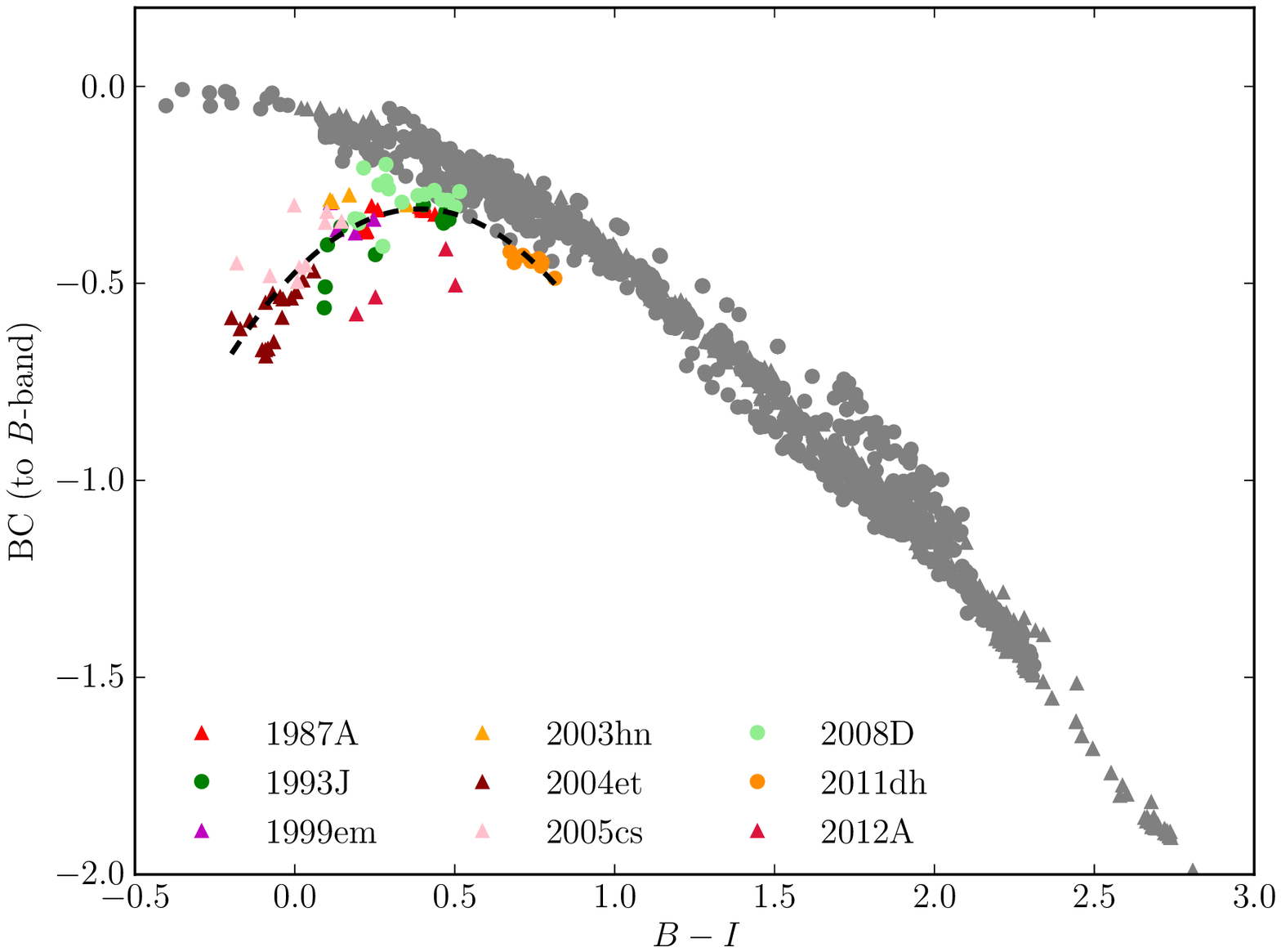}
  \includegraphics[width=\linewidth]{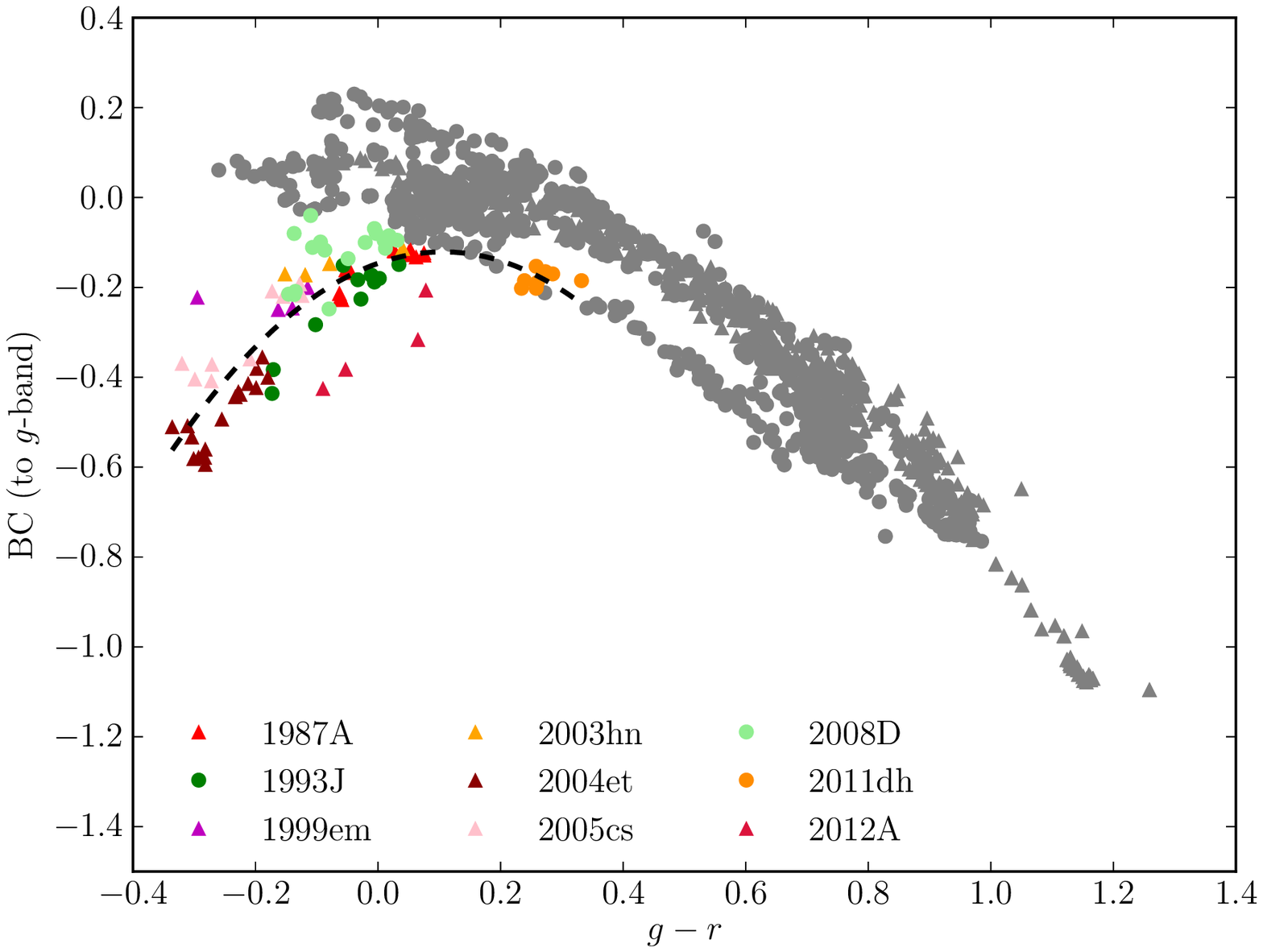}
 \caption{As for Fig.\nobreakspace \ref {fig:bc_all} (grey markers), overlaid with those epochs which exihibit the signature of strong cooling after SBO emission. As is clear these epochs do not occur at unique colours, and as such a separate fit must account for this phase of evolution. A best-fitted second order polynomial is shown for each, fitted to all SNe types (see text).}
 \label{fig:bc_cooling}
\end{figure}

Due to the different treatment of the UV during the cooling phase of SNe evolution from that at later phases, where the BB approximation is more valid than a linear interpolation, we find these epochs require a separate treatment as they are not well described by the parabolas given in Section\nobreakspace \ref {sect:radphase}. This is clearly displayed Fig.\nobreakspace \ref {fig:bc_cooling}, where the epochs over the cooling phase are plotted alongside the data from the radiative and recombination epochs. The fact this cooling phase forms a `branch' in this plot rather than an extension in colour also prompts a separate fit, since the cooling phase occurs over the same optical colours as the later evolution for some SNe.  We follow the same procedure as in Section\nobreakspace \ref {sect:radphase} and fit parabolas to the cooling phase data for each colour. Separate fits for SE~SNe and SNe~II were not done due to the low number of points. The fitted functions are:
\begin{equation}
  \text{BC}_{B,cool} = -0.473 + 0.830 \times (B-I) - 1.064 \times (B-I)^{2}
  \label{eq:bc_cooling_jc}
\end{equation}
\begin{equation}
  \text{BC}_{g,cool} = -0.146 + 0.479 \times (g-r) - 2.257 \times (g-r)^{2}
  \label{eq:bc_cooling_sl}
\end{equation}

The rms value for $B-I$ ($g-i$) is 0.072 (0.078) and the colour range is $-$0.2--0.8 ($-$0.3--0.3). The cooling branch, as expected, is only observed over the bluer colours of SNe evolution, and shows a larger scatter than the later epochs for each colour, which is reflected in the generally larger rms values of the fits given in Section\nobreakspace \ref {sect:otherfits}. The reader's attention is drawn to Section\nobreakspace \ref {sect:uvir_treatment} for a discussion of the UV treatment in this regime.

\subsection{Fits to other colours}
\label{sect:otherfits}

Following \citet{bersten09}, we present all calculated fits for our BC and \pBC{} as tables of coefficients to the polynomials:
\begin{equation}
\text{BC}_{x} = \displaystyle\sum_{i=0}^{2} c_i(x-y)^i
\label{eq:bc_poly}
\end{equation}
\begin{equation}
p\text{BC}_{x} = \displaystyle\sum_{i=0}^{2} c_i(x-y)^i
\label{eq:pbc_poly}
\end{equation}
where BC$_{x}$ and \pBC{}$_{x}$ are the bolometric and pseudo-bolometric corrections to filter $x$, based on colour $x-y$. The coefficients are presented in Tables\nobreakspace \ref {tab:se_params} and\nobreakspace  \ref {tab:ii_params} for SE~SNe and SNe~II respectively. The parameters for the BC appropriate during the cooling phase are provided in Table\nobreakspace \ref {tab:cooling_params}, note these are appropriate for both SE~SNe and SNe~II, as we neglect to divide the sample by type during this phase due to the small numbers involved. Also given are the colour ranges over which the fitted data extend and rms values of the fits in magnitudes.

Note that the data used to produce the $g-i$ fits were corrected for the systematic offset found when estimating $i$-band fluxes from a linear interpolation of the SED, see Appendix\nobreakspace \ref {sect:extractsloan} for more details. However, it was found that the $g-i$ relation has the smallest intrinsic scatter of any colours investigated here, and a fit to this colour should be reassessed once a data set of SNe observed in Sloan filters with good UV/NIR coverage exists.

Fits to $R-I$ and $r-i$ were calculated, but the scatter about these fits was rather larger than the fits presented here, and as such are not included in Tables\nobreakspace  \ref {tab:se_params} to\nobreakspace  \ref {tab:cooling_params} . The larger scatter is probably due to both of these pairs of filters failing to characterise the peak of the SED at any epoch. As such, a given value for either of these colours has a large uncertainty on the strength of the peak of the SED, where the majority of the flux is emitted, and thus a large uncertainty on the BC (or \pBC).

\begin{table*}[]
\caption{Fit parameters for SE~SNe. We indicate in bold the fits with the smallest dispersions (see text).}
\begin{center}
\begin{tabular}{ccc@{\hskip 1.1cm}rrrr@{\hskip 1.1cm}rrrr}

 &   &   &  \multicolumn{ 4}{c}{{\bf BC}} & \multicolumn{ 4}{c}{{\bf \pBC}} \\ 
$x$ & $y$ & $x-y$ range &{$c_0$} & {$c_1$} & {$c_2$} &  {rms}&{$c_0$} & {$c_1$} & {$c_2$} & {rms} \\ 
\hline
$B$ & $V$ & 0.0--1.3  & $-0.083$ & $-0.139$ & $-0.691$ & $0.109$ & $+0.076$ & $-0.347$ & $-0.620$ & $0.112$ \\ 
$B$ & $R$ & 0.1--2.0  & $-0.029$ & $-0.302$ & $-0.224$ & $0.069$ & $+0.136$ & $-0.464$ & $-0.181$ & $0.067$ \\ 
$B$ & $I$ & $-$0.4--2.3  & $-0.055$ & $-0.240$ & $-0.154$ & $\mathbf{0.061}$ & $+0.097$ & $-0.354$ & $-0.131$ & $\mathbf{0.064}$ \\ 
$V$ & $R$ & $-$0.2--0.7  & $+0.197$ & $-0.183$ & $-0.419$ & $0.101$ & $+0.299$ & $-0.372$ & $-0.358$ & $0.087$ \\ 
$V$ & $I$ & $-$0.7--1.1  & $+0.213$ & $-0.203$ & $-0.079$ & $0.090$ & $+0.306$ & $-0.283$ & $-0.084$ & $0.072$ \\ 
% $g$ & $i$ & $-$0.7--1.2 & $+0.006$ & $-0.338$ & $-0.255$ & $\mathbf{0.052}$ & $+0.091$ & $-0.454$ & $-0.211$ & $\mathbf{0.045}$ \\ 
$g$ & $i$ & $-$0.8--1.1& $-0.029$ & $-0.404$ & $-0.230$ & $\mathbf{0.060}$ & $+0.051$ & $-0.511$ & $-0.195$ & $\mathbf{0.055}$ \\ 
$g$ & $r$ & $-$0.3--1.0  & $+0.054$ & $-0.195$ & $-0.719$ & $0.076$ & $+0.168$ & $-0.407$ & $-0.608$ & $0.074$ \\ 
\end{tabular}
\end{center}
\label{tab:se_params}
\end{table*}

\begin{table*}[]
\caption{Fit parameters for SNe~II. We indicate in bold the fits with the smallest dispersions (see text).}
\begin{center}
\begin{tabular}{ccc@{\hskip 1.1cm}rrrr@{\hskip 1.1cm}rrrr}

&   &   &  \multicolumn{ 4}{c}{{\bf BC}} & \multicolumn{ 4}{c}{{\bf \pBC}} \\ 
$x$ & $y$ & $x-y$ range &{$c_0$} & {$c_1$} & {$c_2$} &  {rms}&{$c_0$} & {$c_1$} & {$c_2$} & {rms} \\ 
\hline
$B$ & $V$ & 0.0--1.6  & $-0.138$ & $-0.013$ & $-0.649$ & $0.094$ & $+0.058$ & $-0.331$ & $-0.520$ & $0.092$ \\ 
$B$ & $R$ & 0.1--2.5  & $+0.004$ & $-0.303$ & $-0.213$ & $0.037$ & $+0.124$ & $-0.406$ & $-0.191$ & $0.038$ \\ 
$B$ & $I$ & 0.0--2.8  & $+0.004$ & $-0.297$ & $-0.149$ & $\mathbf{0.026}$ & $+0.121$ & $-0.387$ & $-0.131$ & $\mathbf{0.028}$ \\ 
$V$ & $R$ & 0.0--0.9  & $+0.073$ & $+0.902$ & $-1.796$ & $0.050$ & $+0.059$ & $+1.039$ & $-1.958$ & $0.060$ \\ 
$V$ & $I$ & 0.0--1.2  & $+0.057$ & $+0.708$ & $-0.912$ & $0.043$ & $+0.065$ & $+0.744$ & $-0.953$ & $0.053$ \\ 
% $g$ & $i$ & $-$0.6--1.5  & $+0.029$ & $-0.285$ & $-0.336$ & $\mathbf{0.022}$ & $+0.101$ & $-0.400$ & $-0.289$ & $\mathbf{0.025}$ \\ 
$g$ & $i$ & $-$0.5--1.4  & $-0.007$ & $-0.359$ & $-0.336$ & $\mathbf{0.022}$ & $+0.063$ & $-0.497$ & $-0.268$ & $\mathbf{0.024}$ \\ 
$g$ & $r$ & $-$0.2--1.3  & $+0.053$ & $-0.089$ & $-0.736$ & $0.036$ & $+0.165$ & $-0.332$ & $-0.612$ & $0.037$ \\ 
\end{tabular}
\end{center}
\label{tab:ii_params}
\end{table*}

\begin{table*}[]
\caption{Fit parameters for the cooling phase, appropriate for both SNe types. We indicate in bold the fits with the smallest dispersions (see text).}
\begin{center}
\begin{tabular}{ccc@{\hskip 1.1cm}rrrr}

 &   &   &  \multicolumn{ 4}{c}{{\bf BC}} \\ 
$x$ & $y$ & $x-y$ range &{$c_0$} & {$c_1$} & {$c_2$} &  {rms}\\ 
\hline
$B$ & $V$ & $-$0.2--0.5  & $-0.393$ & $+0.786$ & $-2.124$ & $0.089$\\ 
$B$ & $R$ & $-$0.2--0.8  & $-0.463$ & $+0.790$ & $-1.034$ & $0.078$\\ 
$B$ & $I$ & $-$0.2--0.8  & $-0.473$ & $+0.830$ & $-1.064$ & $\mathbf{0.072}$\\ 
$V$ & $R$ & 0.0--0.4  & $-0.719$ & $+4.093$ & $-6.419$ & $0.125$\\ 
$V$ & $I$ & 0.0--0.4  & $-0.610$ & $+2.244$ & $-2.107$ & $0.146$\\ 
% $g$ & $i$ & $-$0.6--0.2& $-0.127$ & $-0.104$ & $-1.599$ & $\mathbf{0.069}$\\ 
$g$ & $i$ & $-$0.7--0.1& $-0.158$ & $-0.459$ & $-1.599$ & $\mathbf{0.069}$\\ 
$g$ & $r$ & $-$0.3--0.3  & $-0.146$ & $+0.479$ & $-2.257$ & $0.078$\\
\end{tabular}
\end{center}
\label{tab:cooling_params}
\end{table*}

\section{Discussion}
\label{sect:discuss}

Our results show that it is possible to obtain the full bolometric flux of a CCSN from two-filter observations through a simple second-order polynomial correction. Here we will discuss aspects of the results in terms of the BC, although they also largely apply to the \pBC{} relation as well (excluding discussion of UV treatment).

We observe differing scatter for the two samples. As mentioned, SNe~II are expected to be a more homogeneous type of explosion, with the large hydrogen-rich envelopes of the progenitors upon explosion meaning continuum-dominated emission occurs throughout the plateau. The expected sphericity (and likely single-star nature) of the events also means viewing angle will introduce little if any scatter in the relations. We see extremely similar evolution across our SN~II sample, even the peculiar SN1987A. The SE~SNe are subject to other factors that could explain the increased scatter we observe in their relations. Firstly, several progenitor channels are proposed and it is likely that a combination produce the SNe we observe. Binarity and rotation of the progenitor and the intrinsic asphericity of the explosions \citep[e.g.][]{maeda02} are all likely to contribute to scatter in the BC across the sample. High energy components (e.g. gamma-ray burst afterglow components) could be expected also to affect the colours of the SNe. For example we see that SN2008D lies somewhat below the general trend in the $g-r$ fit, and to a lesser extent in $B-I$ fit, as shown in Fig.\nobreakspace \ref {fig:bc_all}, although other SNe with high energy components are well described by the fit (e.g.\ SNe 1998bw and 2006aj). The stripped nature also introduces a range of possible evolution time-scales as more highly stripped progenitors will reveal their heavier elements earlier than those retaining more of their envelopes, making their spectra potentially diverge from homogeneous evolution due to the different chemical composition and pre-mixing of the progenitors.

A factor that could affect the evolution of any SN is the CSM into which it is expanding. Although we have ruled out SNe that show strong interaction with their surrounding medium, in reality, all SNe will have some level of interaction that is dictated by density and composition of the CSM; this being linked to the mass loss of the progenitor system in the final stages of its evolution. Again, this may affect the SE~SN sample more markedly than SNe~II, which are expected to have retained the vast majority of their envelopes until explosion.

\subsection{Treatment of the UV/IR}
\label{sect:uvir_treatment}

Some extremely well-observed SNe have observations that show that the bulk of the light is emitted in the near-ultraviolet (NUV) to near-infrared (NIR) regime. The observed wavelength range investigated here stops at 24400\ \AA{} due to a paucity of data in wavelengths redder than this for CCSNe. \citet{ergon13} show that the MIR regime contributes at most few per cent to their UV-MIR light curve of SN2011dh and the contribution diminishes to negligible values beyond these wavelengths ($\sim$1 per cent). There are no mechanisms producing significant sources of flux at long wavelengths in CCSNe over the epochs investigated here \citep[e.g.][]{soderberg10} and as such the treatment of wavelengths longer than the NIR as a Rayleigh-Jeans law is appropriate. 
%However, the flux missed at shorter wavelengths ($<$~3118\ \AA{}) constitutes a significant fraction of the emitted flux of a SN during early epochs prior to peak. Evolution in the UV regime is diverse and is strongly metallicity-dependent \citep{brown09}. The \emph{Swift} data in \citet{pritchard13} show this through the large scatter in their UVC to optical colours, something we do not see to such extremes here when looking at optical-NIR regimes. UV contributions to UV-optical-NIR light curves are at the level of $\sim$10--20 per cent prior to and at peak, falling to a few per cent soon after peak \citep{modjaz09,stritzinger09,ergon13}. More than a week after peak the UV contribution does not evolve strongly with time and remains at the few per cent level.

Wavelengths shorter than $U$-band constitute a significant fraction of the bolometric flux at certain epochs\footnote{We neglect a treatment of very high energy emission since this is insignificant in terms of bolometric luminosity on the time-scales of SN detections.} and this fraction is difficult to quantify for a large sample of SNe due to the inherently diverse behaviour, the prospect of strong, very blue emission occurring after the SBO in certain SNe, and the fact it is not a very well observed wavelength range in CCSNe. The validity of the treatment of the UV used here (a BB extrapolation to zero\ \AA{} during the cooling phase and a linear extrapolation to zero flux at 2000\ \AA{} for epochs of no strong cooling) was tested using UV observations. Eight SNe of the sample presented (2 SNe~II and 6 SE~SNe) have sufficient existing \emph{Swift} data, as presented in \citet{pritchard13}, to test our method. For each SN with UV data, SEDs were constructed using both the method described in Section\nobreakspace \ref {sect:sed} and the following: instead of extrapolating the UV flux (using the UV approximation appropriate to each epoch), \emph{Swift} UV observations are added to our SEDs, having corrected their magnitudes for reddening using the same method as for the optical and NIR filters. The large red leak of the \emph{uvw2} filter \citep[as demonstrated in relation to SN2011dh by][]{ergon13} was evident from a strong excess in some SEDs for this filter. For this reason the \emph{uvw2} filter was only used for SNe 2007uy, 2008ax and 2012A in epochs $< 2$ weeks from detection, when the blue continuum will minimise contamination in \emph{uvw2} from the red leak. Each SED constructed with \emph{Swift} data was tied to 1615\ \AA{} (the blue cut off of \emph{uvw2}) in all cases. The UV luminosities at each epoch were computed in each case via an integration of the wavelengths from 1615\AA{} (2030\AA{} in the linear extrapolation case) to $U$-band. By comparing the UV luminosity results of each method of SED construction, the accuracy of the UV treatments used here was tested.

\begin{figure}
 \centering
  \includegraphics[width=\linewidth]{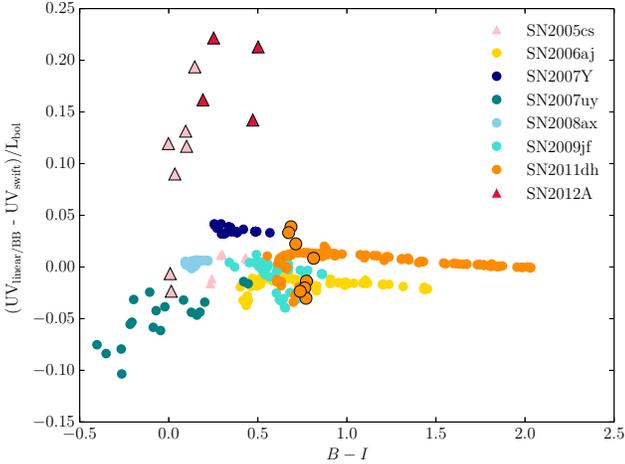}
 \caption{The difference between the UV integrated luminosityies (as a fraction of the bolometric luminosity) between using \emph{Swift} data in SED construction and the UV treatments presented here. Larger, black-edged points represent the epochs where the BB extrapolation method was used for the UV, whereas smaller points represent a linear extrapolation to 2000\AA{} (see Section\nobreakspace \ref {sect:sed_uv}). The methods are consistent within a few per cent except for the case of strong SBO cooling emission for SNe 2005cs and 2012A.}
 \label{fig:swiftuv}
\end{figure}

The results of this test are shown in Fig.\nobreakspace \ref {fig:swiftuv} for the $B-I$ colour. There is generally good agreement between our simple treatments of the UV and when including \emph{Swift} data for the majority of the epochs, with differences in most cases being of the order of a few per cent of the bolometric luminosity. SN2007uy, the SN which shows the largest deviation barring epochs with strong post-SBO cooling (although still $<10$ per cent), has extremely large and uncertain reddening \citep{roy13}. The larger discrepancy between the linear interpolation and the \emph{Swift} data seen for this SN could be indicative of an incorrect reddening value, or reddening law, but it cannot be ruled out that it is intrinsic to the SN. Contributions from wavelengths shorter than 1615\ \AA{} will not contribute much to the bolometric flux except during the cooling phase, when SNe are UV bright. Thus the \emph{Swift} data and, given the good agreement seen, our linear extrapolation method, accounts for the vast majority of the UV flux in a SN.

The cooling branch in this plot, however, displays fairly large discrepancies, even though we are unfortunately limited to 3 SNe (2005cs, 2011dh and 2012A) that have contemporaneous UV-optical-NIR data over the cooling phase. As may be expected from its relatively modest cooling phase, SN2011dh exhibits the best agreement, with even the earliest epochs (\simlt{}2~day after explosion) discrepant by less than 5 per cent at all epochs. The BB treatment of 2005cs and 2012A, both of type II-P, appears to overestimate the UV luminosity at early epochs by 10--20 per cent. An explanation that may account for some of this discrepancy is that the \emph{Swift} SED is tied to zero flux at 1615\AA{} (the limit of the UV integration), whereas the BB will obviously be at some positive flux value. This `cutting-off' of the \emph{Swift} SED is an under estimation of the flux, especially in these extremely blue phases, but a lack of data at shorter wavelengths necessitates this treatment. Two very early epochs of the evolution of SN~2005cs are well matched by the BB treatment however, and it may be that the later cooling phase epochs are falling from the BB approximation quicker than expected. The intrinsically heterogeneous nature of this cooling phase is evident in the large scatter observed during these epochs (Fig.\nobreakspace \ref {fig:bc_cooling}) and we must also add the caveat that our simple UV treatment may be discrepant at the 10--20 per cent level. This discrepancy, however, appears only evident in SNe~II-P and at the very early epochs. An increase in sample size is desired to further quantify this and thus improve upon the UV treatment at these epochs. Given this, for events where UV data exist which is indicative of post-SBO cooling emission, it is advisable to use the \pBC{} and add the UV contribution directly from observations.

\subsection{Time-scales of validity}
\label{sect:epochrange}

\begin{figure}
 \centering
 \includegraphics[width=\linewidth]{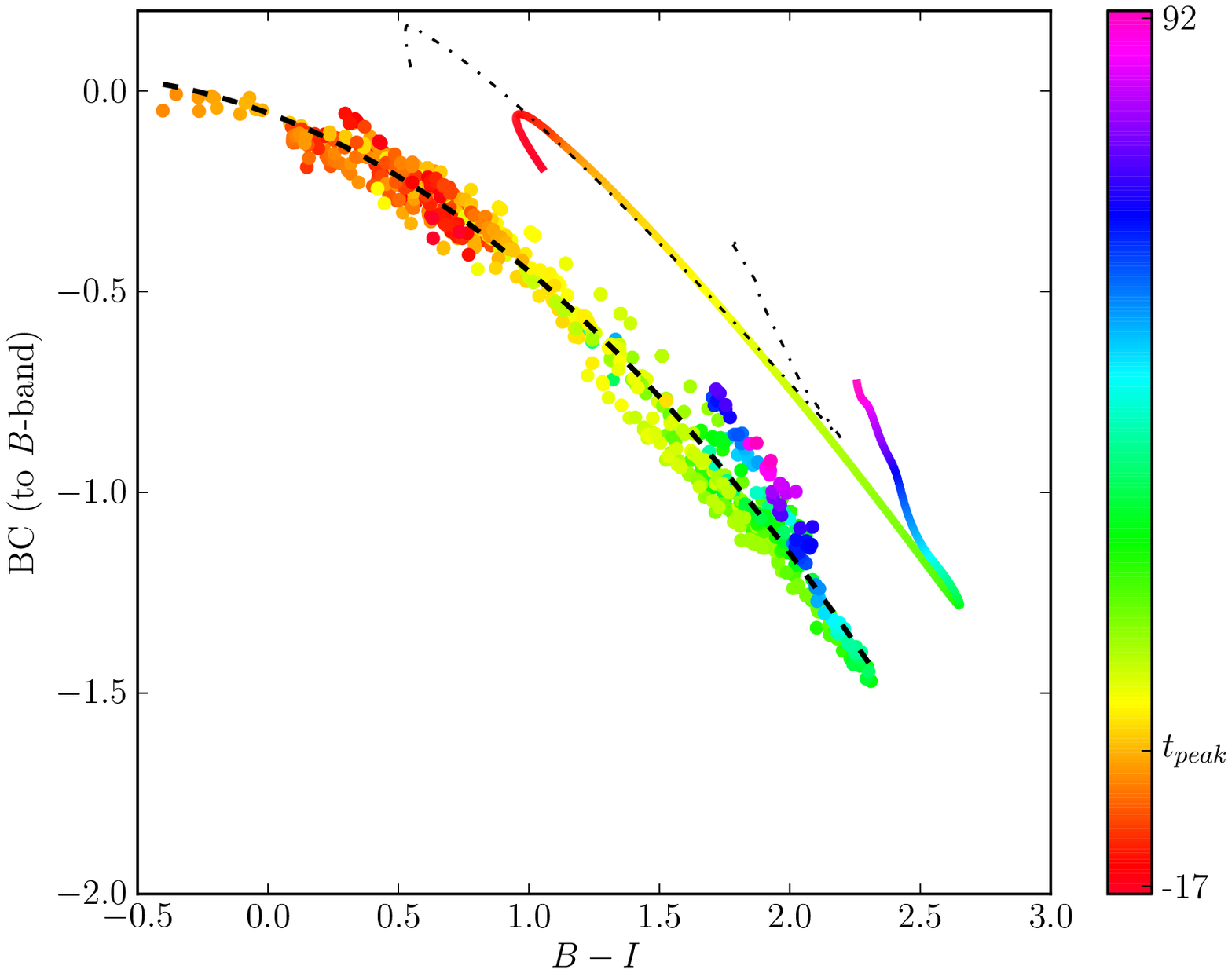}
 \includegraphics[width=\linewidth]{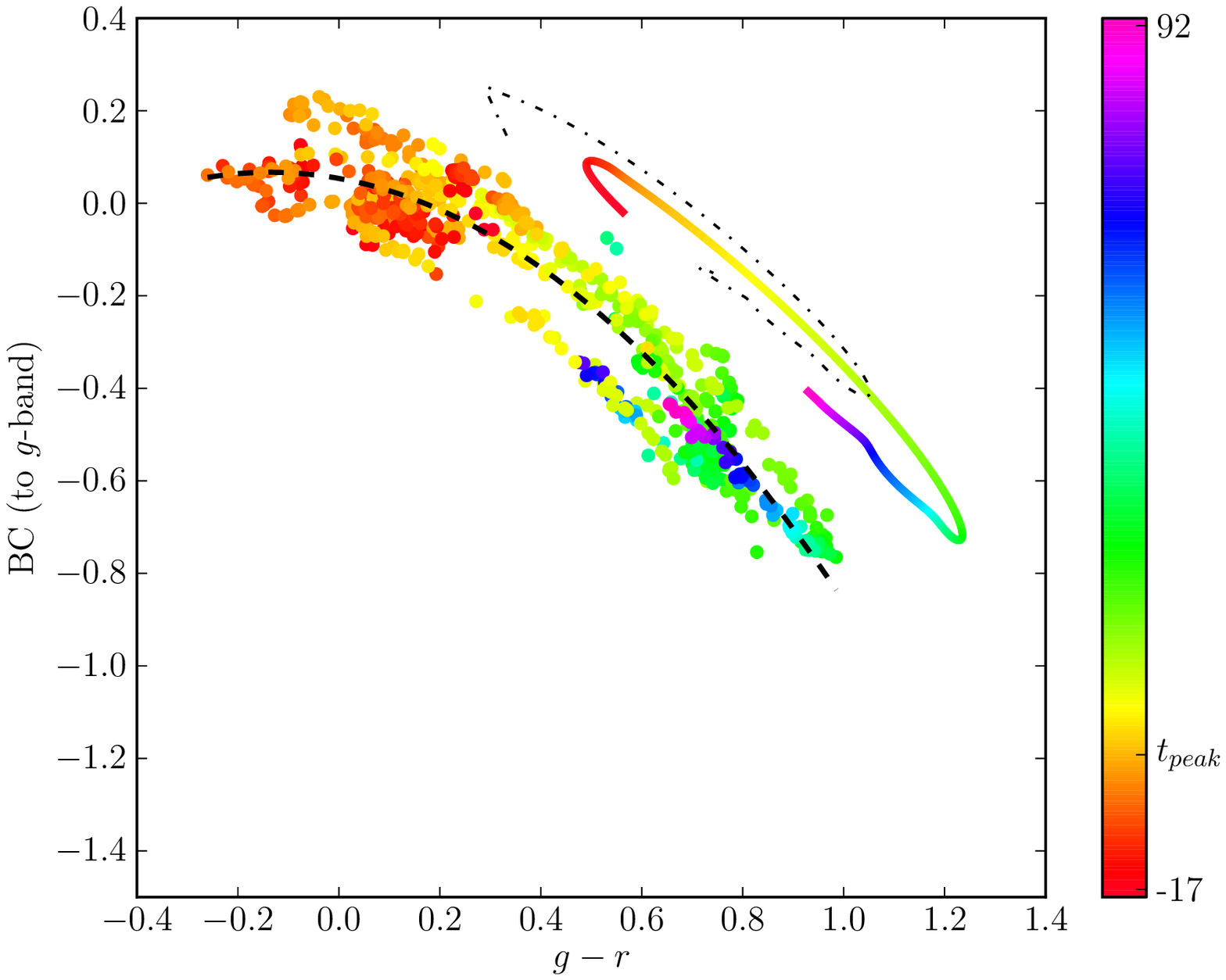}
 \caption{The BC for SE~SNe, colour-coded to show evolution with time. The colour bar indicates the phase with respect to the $V$-band peak ($t_{peak}$), where the SN evolution has been stretched such that \dmf{} $= 0.758$ (see Section\nobreakspace \ref {sect:fluxcont}). Epochs shown \emph{do not} include those exhibiting signatures of strong cooling after SBO. To illustrate the typical movement of a SN in this plot, the polynomial-smoothed evolution of SN2011dh is plotted offset from the data. The effect of reddening is shown by re-analysing SN2011dh with an increase in $E(B-V)$ of 0.2 (black dot-dash line). Equation\nobreakspace \textup {(\ref {eq:bc_se_sl})} (top) and Eq.\nobreakspace \textup {(\ref {eq:bc_se_jc})} (bottom) are also plotted for each filter set (thick black dashed line).}
 \label{fig:epoch_se}
\end{figure}

It is important to determine over what epoch range these relations are valid for each sample. Fig.\nobreakspace \ref {fig:epoch_se} shows the evolution in time of the SE~SNe in the BC plot. The intrinsic scatter about the fits does not change dramatically with epoch and a largely coherent evolution from top-left to bottom-right in each plot is observed, with very late-time data beginning to move top-left again. The duration of validity after the peak is tied to the time-scale of evolution of the SNe. We have normalised our SE~SN evolution by making use of the \dmf{} value for each SN (Section\nobreakspace \ref {sect:fluxcont}). When evolution is normalised by this factor, the two SNe with the data furthest past peak are SN2011dh ($\sim$93 days) and SN2007gr ($\sim$75 days). The data representing these late epochs are clearly visible in Fig.\nobreakspace \ref {fig:epoch_se}, with the evolution of SN2011dh explicitly shown offset from the data -- both show a trend towards moving above (below) the $B-I$ ($g-r$) fit at late epochs. Data covering \UK{} for SN2007gr actually extend to roughly 120 days after optical peak. These data are not included as they diverge from the correlation, as appears to be happening for SN2011dh. SN2011dh and SN2007gr are at the higher end of the \dmf{} range (0.968 and 0.861, respectively) and may be considered to give a good limit for the range of validity of this fit. The data presented show the corrections for SE~SNe to be valid from shortly after explosion (earliest data are $\sim$2 days post-explosion) to $\sim$50 days past peak, and potentially further, although we are limited to analysing only two SNe.

\begin{figure}
 \centering
 \includegraphics[width=\linewidth]{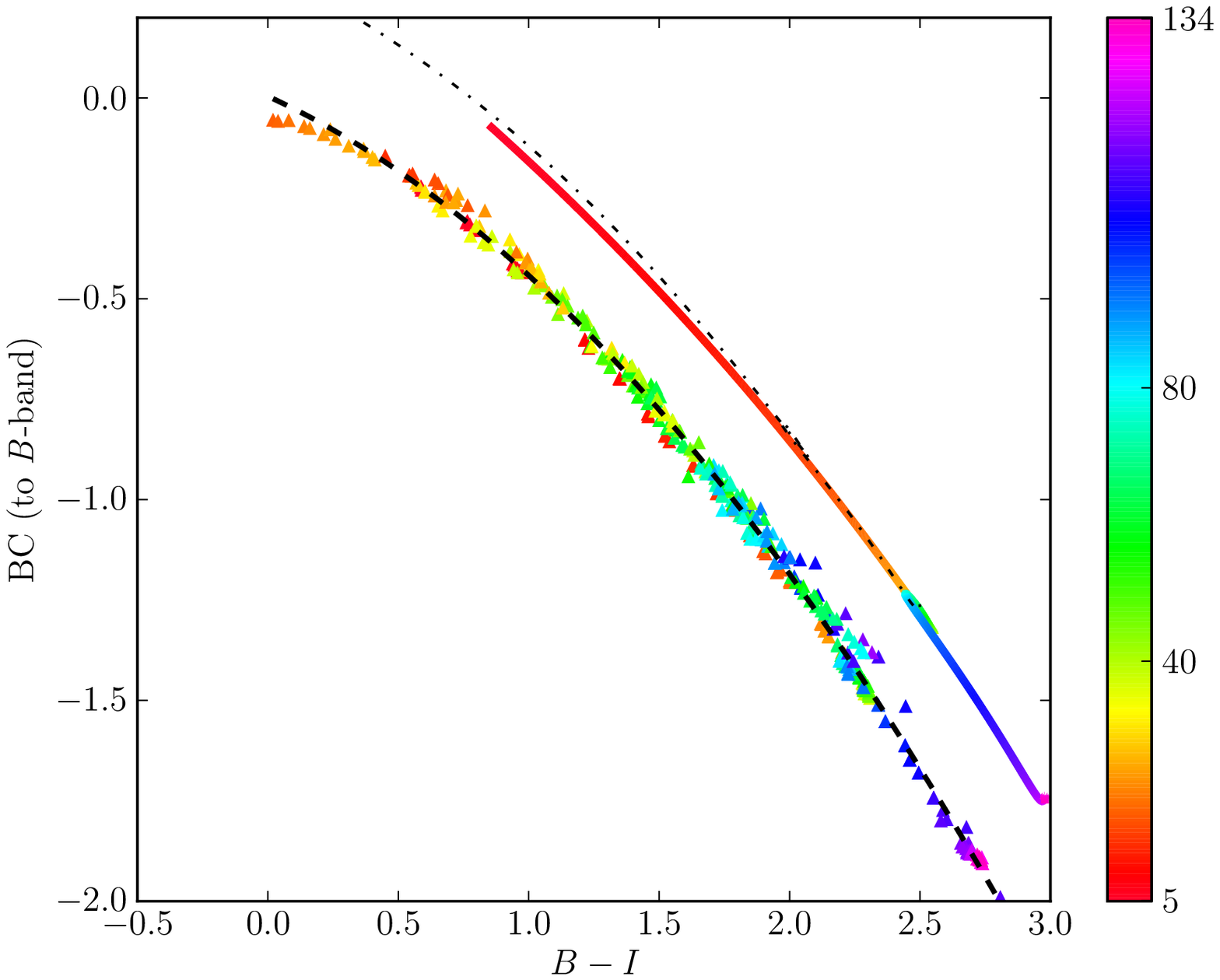}
 \includegraphics[width=\linewidth]{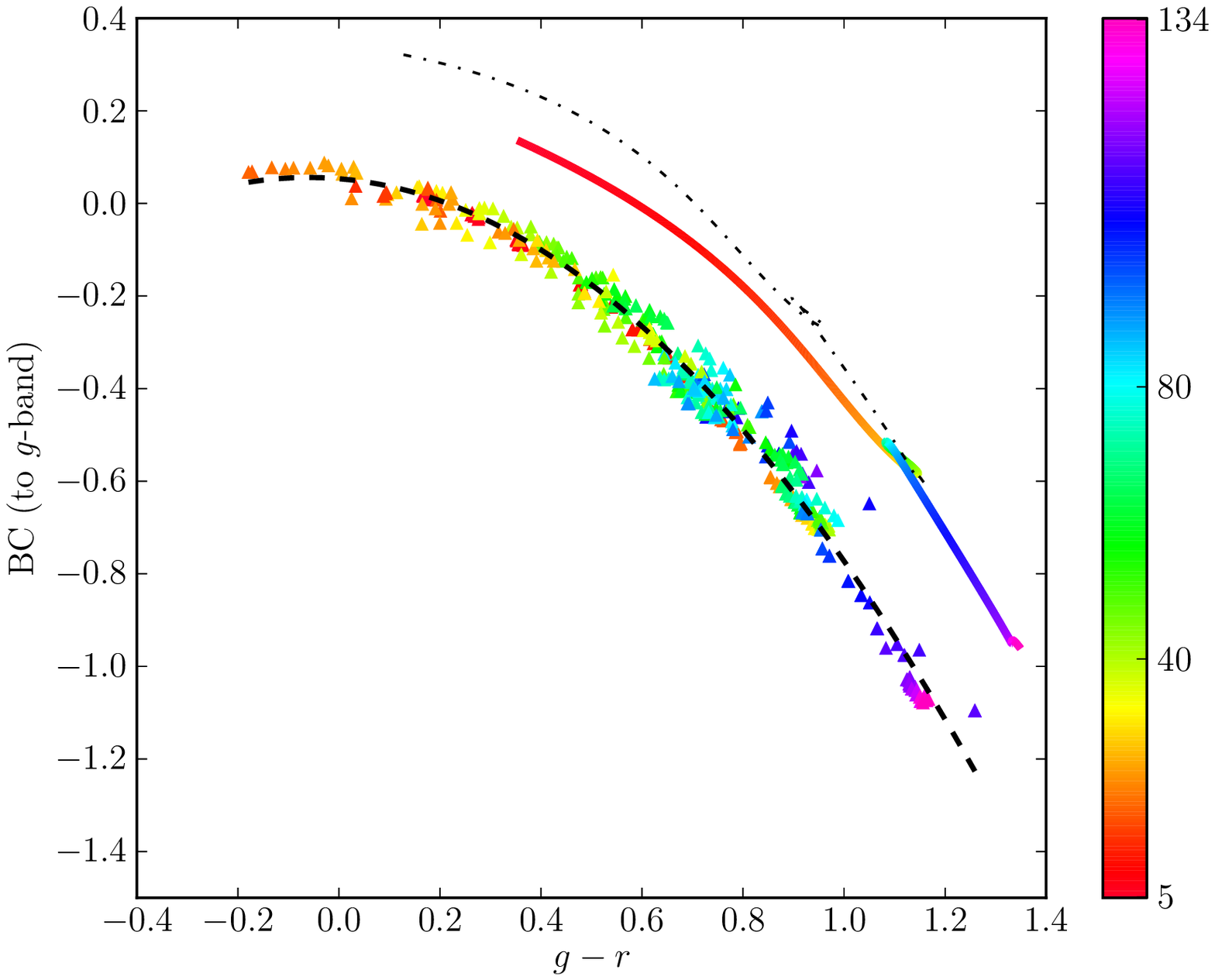}
 \caption{The BC for SNe~II, colour-coded to show evolution with time. The colour bar indicates the phase with respect to the explosion date. Epochs shown \emph{do not} include those exhibiting signatures of strong cooling after SBO. To illustrate the typical movement of a SN in this plot, the polynomial-smoothed evolution of SN1987A is plotted offset from the data. The effect of reddening is shown by re-analysing SN1987A with an increase in $E(B-V)$ of 0.2 (black dot-dash line). Equation\nobreakspace \textup {(\ref {eq:bc_ii_sl})} (top) and Eq.\nobreakspace \textup {(\ref {eq:bc_ii_jc})} (bottom) are also plotted for each filter set (thick black dashed line).}
 \label{fig:epoch_ii}
\end{figure}

Figure\nobreakspace \ref {fig:epoch_ii} shows the evolution of the BC for the SN~II sample, where the colour indicates days from explosion date. For the SN~II sample we see that even very early data (e.g.\ beginning at $\sim$5 days past explosion for SN1987A) have a small dispersion. Evolution in this plot appears to be simpler than the SE~SNe with a smooth transition from top-left to bottom-right. However, SN1987A undergoes a phase of little evolution in colour (and BC) from days $\sim$40--80, with other SNe~II displaying a similar period of inactivity in the plot during the plateau phase. Despite the fact SN1987A also appears to evolve much more rapidly and evolves to much redder colours, as can be seen in Fig.\nobreakspace \ref {fig:fluxvstime}, its evolution is still remarkably consistent with the other objects in the BC plots, and its additional, redder, evolution follows the parabolic fit. The phase range investigated here is broadly over the plateau of SNe~II, after which the deeper layers of the ejecta act to destroy any homogeneous evolution. For example, \citet{inserra12} show optical colours for several SNe~II-P to late times, with diverse behaviour observed after $\sim$120 days (the end of the plateau). This can also be seen in the BCs presented by \citet{bersten09}, where the BC scatter increases dramatically after the end of the plateau. We therefore limit the use of these fits from explosion until the time of transition from plateau to radioactive tail.

It must be stressed however that the use of these fits will primarily be for SNe detected only in the optical regime. As such, there is no knowledge of any UV bright SBO cooling emission, given that the optical colour ranges overlap for the cases of strong and no cooling emission (as shown for the fits in Fig.\nobreakspace \ref {fig:bc_cooling}). Relying only on optical follow up, although vastly increasing the number of SNe with the requisite data, means there is uncertainty in the early light curve.  Hence, although the above described fits \emph{are} valid at early epochs, they are valid only for the case of no strong SBO cooling emission. In the case where unobserved SBO emission is present, the fits will under predict the actual bolometric luminosity. In such cases, use of the cooling phase fits will provide an alternate, plausible, bolometric luminosity in these early epochs by assuming the case of strong SBO cooling emission. This uncertainty can be coupled with previous knowledge of the durations of SBO cooling emission and the type of SN. For example a SN~Ib/c would not be expected to have SBO cooling emission beyond 1--2 days and the cooling fit would over estimate the luminosity at further epochs. Complementary data indicative of SBO cooling emission would warrant the sole use of the cooling phase fit for those epochs, or the use of the \pBC{} and a separate treatment of the UV emission from the available data. The cooling phase fits include data from early after explosion ($\sim$2 days) to the end of the SBO cooling being dominant.

\subsection{Reddening}

An uncertainty when constructing the SEDs is the reddening towards each SN. Although Galactic reddening may be well known, $E(B-V)_{host}$ values are generally less certain. Additional to this, the reddening law for each host is assumed to match that of the Galaxy, an assumption made in the absence of detailed knowledge of the gas and dust properties of the hosts. An increase in assumed $E(B-V)$ will cause a decrease in $B-I$ and $g-r$ (i.e.\ make the SN intrinsically bluer); this will also affect the BC, however. The BC becomes more positive with increasing $E(B-V)$ as the $g$ (or $B$)-band value increases more rapidly than the bolometric magnitude for a given change in $E(B-V)$. Combining these effects means that the SNe actually move somewhat along the fits when reddening is varied. This effect is plotted in Figs.\nobreakspace \ref {fig:epoch_se} and\nobreakspace  \ref {fig:epoch_ii} via an artificial increase of 0.2 in $E(B-V)$ for the offset SN in each plot. Moderate reddening uncertainties do not affect the actual value of the fits drastically, although clearly an accurate reddening value is desired when using the fits, to ensure the SN's true colour for a given epoch is measured (and consequently the correct value for the BC is used).

\section{Summary}

We have presented simple fits making it possible to easily obtain accurate estimates of the bolometric light curves of any CCSN given only two filter observations. We have presented both Johnson-Cousins and Sloan colour corrections and shown our method for determining Sloan magnitudes is robust for $g$ and $r$, and accounted for the systematic offset present in $i$-band determinations. Fits to $B-I$ and $g-r$ are presented as the best fits to each filter set. The SE~SNe corrections (Eqs.\nobreakspace \textup {(\ref {eq:bc_se_jc})} and\nobreakspace  \textup {(\ref {eq:bc_se_sl})}) are constrained in the colour ranges $-0.4 < B-I < 2.3$ (rms 0.061~mag) and $-0.3 < g-r < 1.0$ (0.076~mag). The SNe~II corrections (Eqs.\nobreakspace \textup {(\ref {eq:bc_ii_jc})} and\nobreakspace  \textup {(\ref {eq:bc_ii_sl})}) hold for  $0.0 < B-I < 2.8$ (0.026~mag) and $-0.2 < g-r < 1.3$ (0.036~mag). Corrections for other optical colours are presented; these corrections are valid over the radiatively/recombination powered, photospheric epochs of CCSN evolution. Evolution during epochs that show cooling following SBO emission are fitted separately (Eqs.\nobreakspace \textup {(\ref {eq:bc_cooling_jc})} and\nobreakspace  \textup {(\ref {eq:bc_cooling_sl})}) and are subject to larger uncertainties. Given the diversity and uncertainty of UV evolution, separate pseudo-bolometric fits are given where no treatment of the UV regime is made.

The bolometric corrections presented here will allow current and future SNe surveys, where the sheer number of detected events prevents intense monitoring of the large majority of SNe, to accurately and easily use their optical detections to obtain estimates of the bolometric light curves of CCSNe of all types, essential for modelling of such events.

\section*{Acknowledgements}

The authors wish to thank the anonymous referee for helpful comments and suggestions, which led to a significantly improved paper. This research has made use of the NASA/IPAC Extragalactic Database (NED) which is operated by the Jet Propulsion Laboratory, California Institute of Technology, under contract with the National Aeronautics and Space Administration. The Weizmann interactive supernova data repository (www.weizmann.ac.il/astrophysics/wiserep) was used to obtain SN spectra. JDL acknowledges the UK Science and Technology Facilities Council for research studentship support.

% ------------------Bibliography-----------------------------------------------
% bibliography
\newcommand{\araa}{ARA\&A}   \newcommand{\aap}{A\&A}
\newcommand{\aj}{AJ}         \newcommand{\apj}{ApJ}
\newcommand{\apjl}{ApJ}      \newcommand{\apjs}{ApJS}
\newcommand{\mnras}{MNRAS}   \newcommand{\nat}{Nature}
\newcommand{\pasj}{PASJ}     \newcommand{\pasp}{PASP}
\newcommand{\procspie}{Proc.\ SPIE} \newcommand{\physrep}{Phys. Rep.}
\newcommand{\apss}{APSS}
\newcommand{\solphys}{Sol. Phys.}
\newcommand{\actaa}{Acta Astronom}
\newcommand{\aaps}{A\&A Supp}
\newcommand{\iaucirc}{IAU Circular}
\bibliographystyle{mn2e}
\bibliography{/home/jdl/references}

\appendix

\section{Extracting Sloan magnitudes from Johnson-Cousins SEDs}
\label{sect:extractsloan}

It is desirable, particularly given the impending prevalence of their use for large-scale surveys, to present bolometric corrections using Sloan filters. The SN sample here, however, is literature-based, and as such is predominantly observed in Johnson-Cousins filters. Sloan magnitudes were derived for these SNe by extracting fluxes from the SEDs constructed with Johnson-Cousins measurements at the corresponding \leff{} of the filters. Large deviations from a linear interpolation between neighbouring filters are not expected since the widths of the filters ($\sim$2000\ \AA{}) are much larger than individual spectral features and there is a large amount of overlap between Sloan filters and their neighbouring Johnson-Cousins counterparts. However, it may be that some systematic error could be introduced by this method, particularly when one considers the effect of the extremely strong \Ha{} emission line in SNe~II, or that the strong Ca II NIR triplet absorption borders on the $I$-band. These deviations could make the linear interpolation a poor estimate of the true flux in some specific SN types or at specific epochs.

To test the linear interpolation method we have collated all spectra in WISeREP of our SNe sample that cover a sufficient wavelength range (i.e.\ completely cover the transmission profile of at least one of the $gri$ filters and the appropriate neighbouring Johnson-Cousins filters) and are at similar epochs to those when the photometric SEDs are constructed (see Table\nobreakspace \ref {tab:sn_sample}). Synthetic photometry was performed on the de-reddened spectra (using the $R=3.1$ curve of \citealt{fitzpatrick99}) via:
\begin{equation}
 F_x = \frac{\int  \! T_x(\lambda)f(\lambda)\lambda \text{d}\lambda}{\int  \! T_x(\lambda)\lambda \text{d}\lambda}
\end{equation}
to obtain the flux in filter $x$, based on spectral flux of the SN, $f(\lambda)$, and the transmission profile of filter $x$, $T_x(\lambda)$. This was done for each of the \emph{griBVRI} filters where spectral coverage allowed. After accounting for the filter zeropoints, a directly measured value of the magnitude in each the filters was obtained. Neighbouring Johnson-Cousins filters were interpolated linearly in flux between their \leff{} values such that a flux (subsequently converted to a magnitude) for the appropriate Sloan filter could be made at its \leff{} (e.g. flux values for $B$ and $V$ were interpolated to obtain an estimate for the $g$-band).  

\begin{figure}
 \centering
 \includegraphics[width=\linewidth]{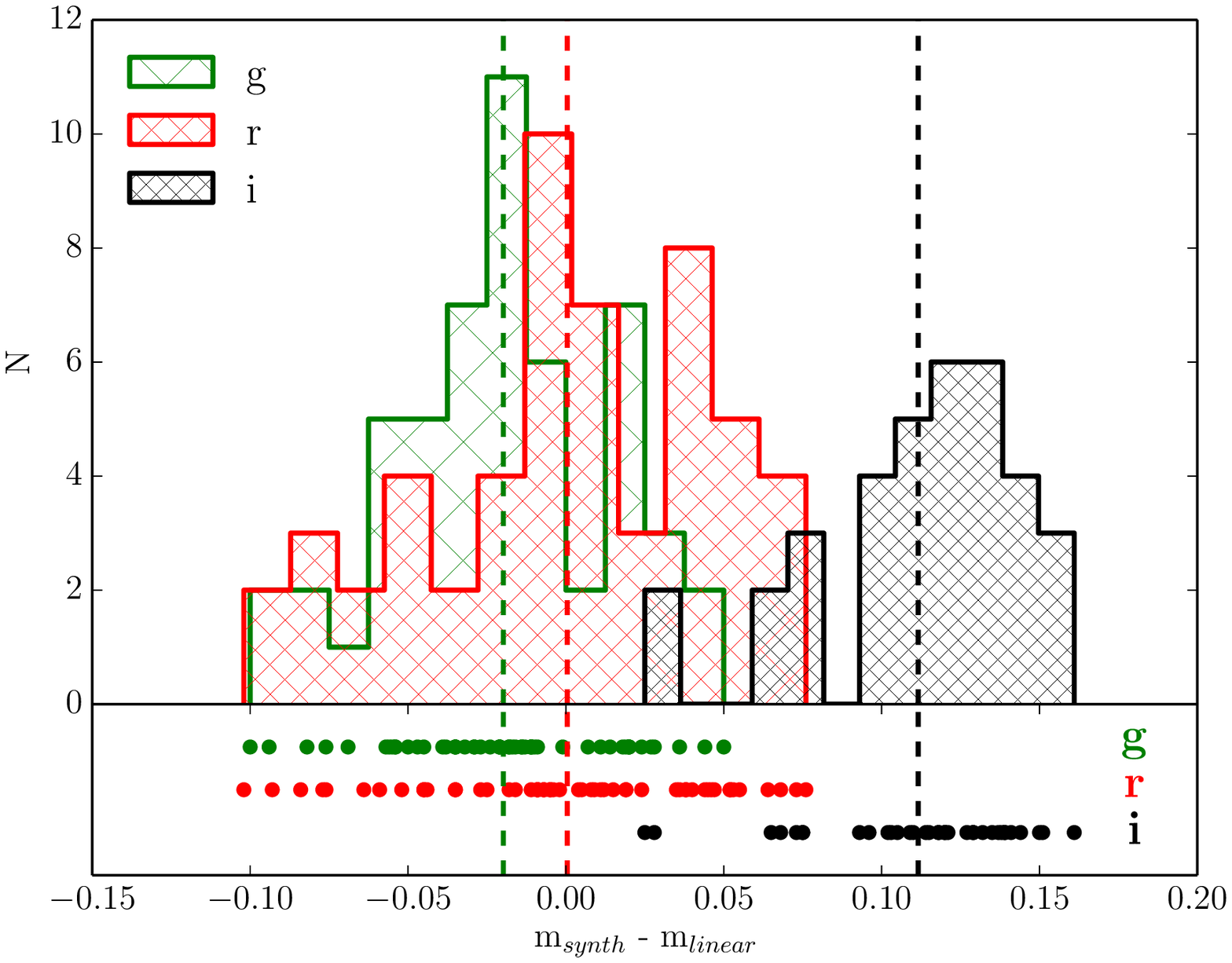}
 \caption{Residuals found between obtaining Sloan synthetic magnitudes from spectra (m$_{synth}$) and those using a linear interpolation from neighbouring Johnson-Cousins filters (m$_{linear}$) for all spectra of the SN sample during the epochs where SEDs have been constructed in the main investigation. The mean values for the three distributions are marked as the vertical dashed lines.}
 \label{fig:synth_gri}
\end{figure}

The comparison of the Sloan magnitudes found via direct synthetic photometry (m$_{synth}$) and using the linear interpolation method (m$_{linear}$) is presented in Fig.\nobreakspace \ref {fig:synth_gri}. Firstly, it was found that there was no dependency on SN type or epoch for the values of $\text{m}_{synth}-\text{m}_{linear}$. Given this, the $g$ and $r$ mean offsets fall extremely close to zero, with values of -0.019 (rms 0.034) and 0.0004 (0.044)~mag respectively. We can therefore say our method is not introducing any systematic offset for quantities reliant on these filters and as such consider the BC and \pBC{} fits to $g$ against $g-r$ colour to be valid. 

However, it is true we see some systematic offset in the measurements of the $i$-band, with the mean being 0.111 (0.032)~mag, despite the Sloan $i$ and Cousins $I$-band transmission curves overlapping heavily. This overestimation of $i$-band flux from the linear interpolation method suggests the $g-i$ colours derived from SED interpolations are likely to be redder than the intrinsic $g-i$. For this reason we increase the magnitude values obtained for $i$ in our linear interpolation method by the mean offset (0.111~mag), to give estimates that will be free from this systematic offset. These corrected $i$ values were used when computing fits to $g-i$ that are presented in Tables\nobreakspace  \ref {tab:se_params} to\nobreakspace  \ref {tab:cooling_params} .

The difference in the BC (and \pBC{}) due to this change to $i$ values for a given $g-i$, when compared to calculating the fits without accounting for the offset, is 0.103~mag for the reddest SE~SNe colour (i.e. $g-i = 1.1$), with differences reaching 0.198~mag for the reddest SNe~II colours ($g-i = 1.4$). The differences are less for typical SN colours and approaching zero for bluer colours ($g-i < -0.5$). Similarly for the cooling phase fit, differences for the bluest colours ($g-i < -0.5$) reach $0.182$~mag, with the average difference being $0.078$~mag. 

An investigation into this systematic offset in $i$ showed the choice of \leff{} to be mainly responsible. There is no knowledge of the spectral features of a SN based on photometric data, above that which can be discerned from colours. As such we are limited, as has been done in similar previous work, to using a single \leff{} for each filter (taken from the literature, or based on the transmission profile), when in reality this should change according to the shape of the underlying spectrum being observed, as in:

\begin{equation}
 \lambda_{eff} = \frac{\int  \! T_x(\lambda)f(\lambda)\lambda^2 \text{d}\lambda}{\int  \! T_x(\lambda)f(\lambda)\lambda \text{d}\lambda}
\end{equation}
for photon-counting devices \citep[e.g.][]{bessell12}.

It was found that the \leff{} for $I$-band when using this equation was particularly susceptible to large deviations compared to the value we used when constructing SEDs, whereas filters $BVR$ were more stable. This is most likely due to the $I$-band being consistently on the `tail' of the flux distribution, coupled with strong absorption due to the calcium triplet bordering on the red edge of the transmission curve. These effects weight the \leff{} to lower values, although no significant dependence on epoch was found either in the change of \leff{} or the value of the offset in $\text{m}_{synth}-\text{m}_{linear}$ for $i$. This change in \leff{} obviously impacts on the gradient of the linear interpolation between $R$ and $I$, and hence on the estimate of $i$. When accounting for the changing \leff{}, the mean systematic offset in $i$ was more than halved, whilst $g$ and $r$ remained very close to zero. However, we take no account of changing \leff{} in our SED construction since we have no a priori information on spectral shape from photometry. Hence we opt not to include it here in our test of the method, but present these findings to highlight the potential uncertainties in the choice of \leff{} when constructing SEDs from photometric data for SNe.

We therefore conclude the determination of Sloan magnitudes derived from our method of linearly interpolating the SEDs constructed with Johnson-Cousins filters are robust \emph{for the cases of $g$- and $r$-bands}, however \emph{there is a systematic offset in the $i$-band magnitude determination} from the two methods. We neglect to correct our derived $g$- and $r$-band magnitudes, given the mean difference is very close to zero in both cases, but do correct our $i$-band measurements by the mean offset before calculating fits based on this filter. It is obviously desirable to reassess the fits to these filters once an appreciable data set of SNe observed directly in Sloan filters with good NIR (and ideally UV) coverage exists.

\section{SNe 1987A and 2009jf: test cases}
\label{sect:09jf}

In order to practically test the reliability of the fits, we recover the bolometric light curves of one SN in each sample.

\subsection{SN1987A}

Being the best observed SN to date, SN1987A represents a good test of the method. Here we reconstruct the bolometric light curve of SN1987A using the optical photometry of \citet{menzies87}. Extinction was corrected for using $E(B-V) = 0.17$ and the procedure described in Section\nobreakspace \ref {sect:sed_opt}. Two fits with differing rms values were used: $B-V$ and $B-I$. Equation\nobreakspace \textup {(\ref {eq:bc_poly})} was used with the appropriate fit parameters (the first 2.1~days were calculated with the cooling phase fits given in Table\nobreakspace \ref {tab:cooling_params}, subsequent epochs used the appropriate BC in Table\nobreakspace \ref {tab:ii_params}) to turn each of these colours into a bolometric magnitude using Eq.\nobreakspace \textup {(\ref {eq:bc})}. Using a distance modulus of $\mu = 18.46$, this was converted to an absolute magnitude and then to a bolometric luminosity via Eq.\nobreakspace \textup {(\ref {eq:bolo})}. Alongside this we also make use of the BC presented in \citet{bersten09} to reproduce their bolometric light curve for SN1987A. Finally we also use the data of \citet{suntzeff90} for the UV-optical-IR observed bolometric light curve of SN1987A for comparison.

\begin{figure}
 \centering
 \includegraphics[width=\linewidth]{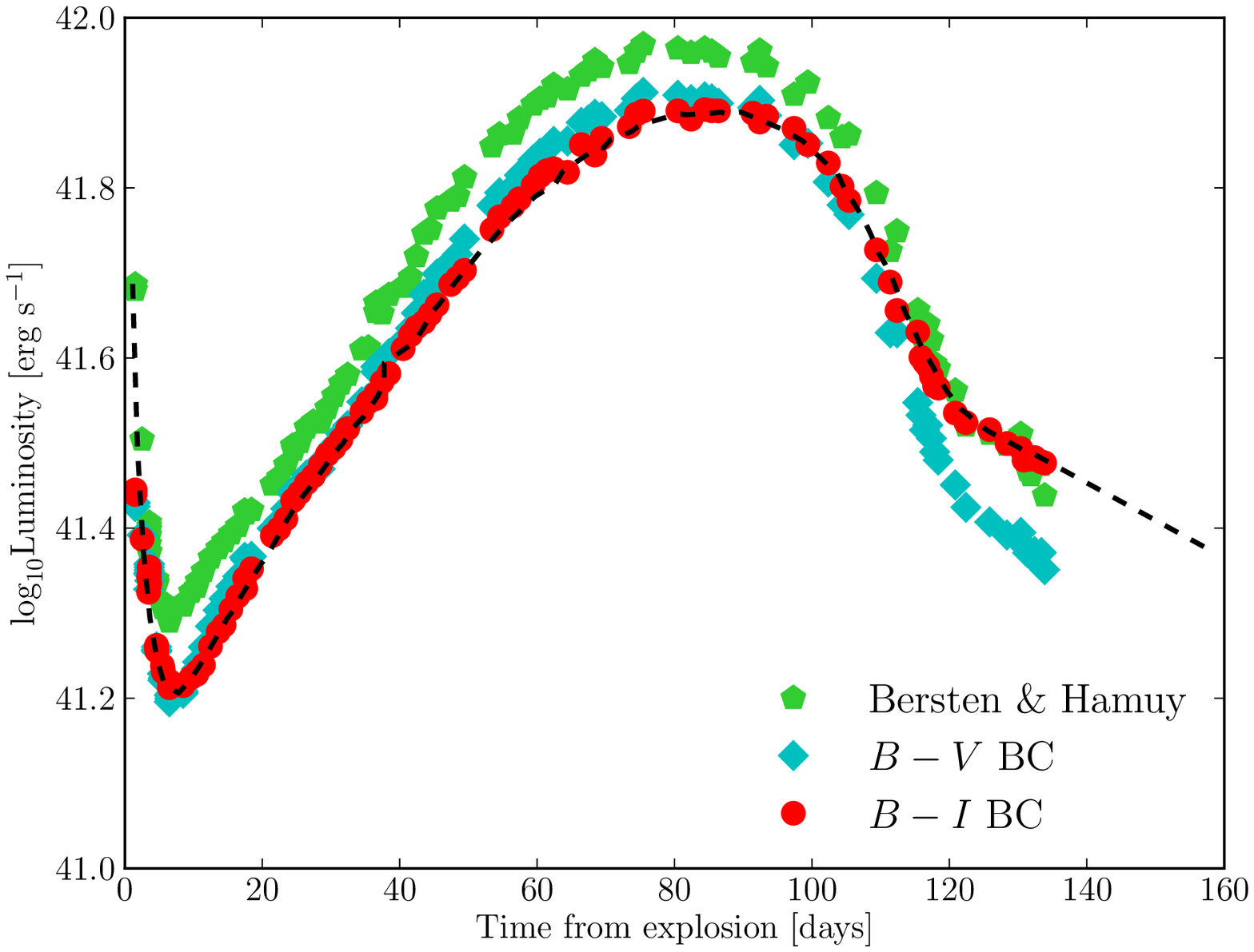}
 \caption{The bolometric light curve of SN1987A constructed from optical colours using the $B-I$ and $B-V$ corrections presented here, and the fit of \citet{bersten09}. Also plotted is the observed bolometric light curve of \citet{suntzeff90} (black dashed line).}
 \label{fig:87A_bolo}
\end{figure}

The four bolometric light curves of SN1987A are shown in Fig.\nobreakspace \ref {fig:87A_bolo}. First to note is the good agreement between all four methods, and particularly the agreement between the BC methods and the observed bolometric light curve. The extremely early data $<$1~day are slightly underestimated, and we again stress here the caveat that emission from cooling post-SBO is subject to larger uncertainties when using these fits. After this, including the tail of the cooling phase, we observe excellent agreement over the rest of the evolution. Although our IR extends nominally to infinity, the observed data stop in the MIR. The excellent agreement between our $B-I$-derived bolometric luminosity and the observed bolometric light curve thus suggests that these very long wavelengths make little difference to the bolometric flux, and certainly not at a level to affect derived parameters from modelling.

Happily enough, we also note that when using a fit with larger scatter (i.e.\ $B-V$), we recover a bolometric light curve consistent to that produced by the $B-I$ fit and, importantly, the observed light curve. Some deviation is observed $>$110 days and this is most likely a result of the two relatively blue filters used in this colour not tracing the evolution of the IR particularly well.

\subsection{SN2009jf}

SN2009jf has good coverage over rise, peak and decline in \emph{UBVRIJHK} filters (used to construct the SED in Section\nobreakspace \ref {sect:sed_opt}). However, alongside these data there exists a set of well-calibrated Sloan observations \citep{valenti11}, including good $g$- and $r$-band coverage from which to test the Sloan fit -- both in terms of the actual fits as well as using the SED interpolations to extract Sloan magnitudes. The $g$ and $r$ data were used to construct the pseudo- and full bolometric light curve of SN2009jf.

Initially the $g$ and $r$ data of \citet{valenti11} were corrected for reddening, which was done in the manner of Section\nobreakspace \ref {sect:sed} with $E(B-V)_{tot} = 0.117$~mag. For epochs of simultaneous $g$- and $r$-band observations, $g-r$ values (corrected for reddening) were fed into Eqs.\nobreakspace \textup {(\ref {eq:bc_poly})} and\nobreakspace  \textup {(\ref {eq:pbc_poly})} using the parameters in Table\nobreakspace \ref {tab:se_params} to obtain values of the BC and \pBC{}, these were converted into apparent bolometric and pseudo-bolometric magnitudes via Eq.\nobreakspace \textup {(\ref {eq:bc})} using the $g$ magnitudes. The distance modulus ($\mu = 32.65$) converted these to an absolute magnitudes, and finally Eq.\nobreakspace \textup {(\ref {eq:bolo})} was used to convert this to a luminosity.

The SED (constructed as in Section\nobreakspace \ref {sect:sed}) was integrated from 2000\ \AA{} to infinity and $U$-band to infinity to obtain the observed bolometric and pseudo-bolometric light curves from \UK{} photometric data. 

\begin{figure}
 \centering
 \includegraphics[width=\linewidth]{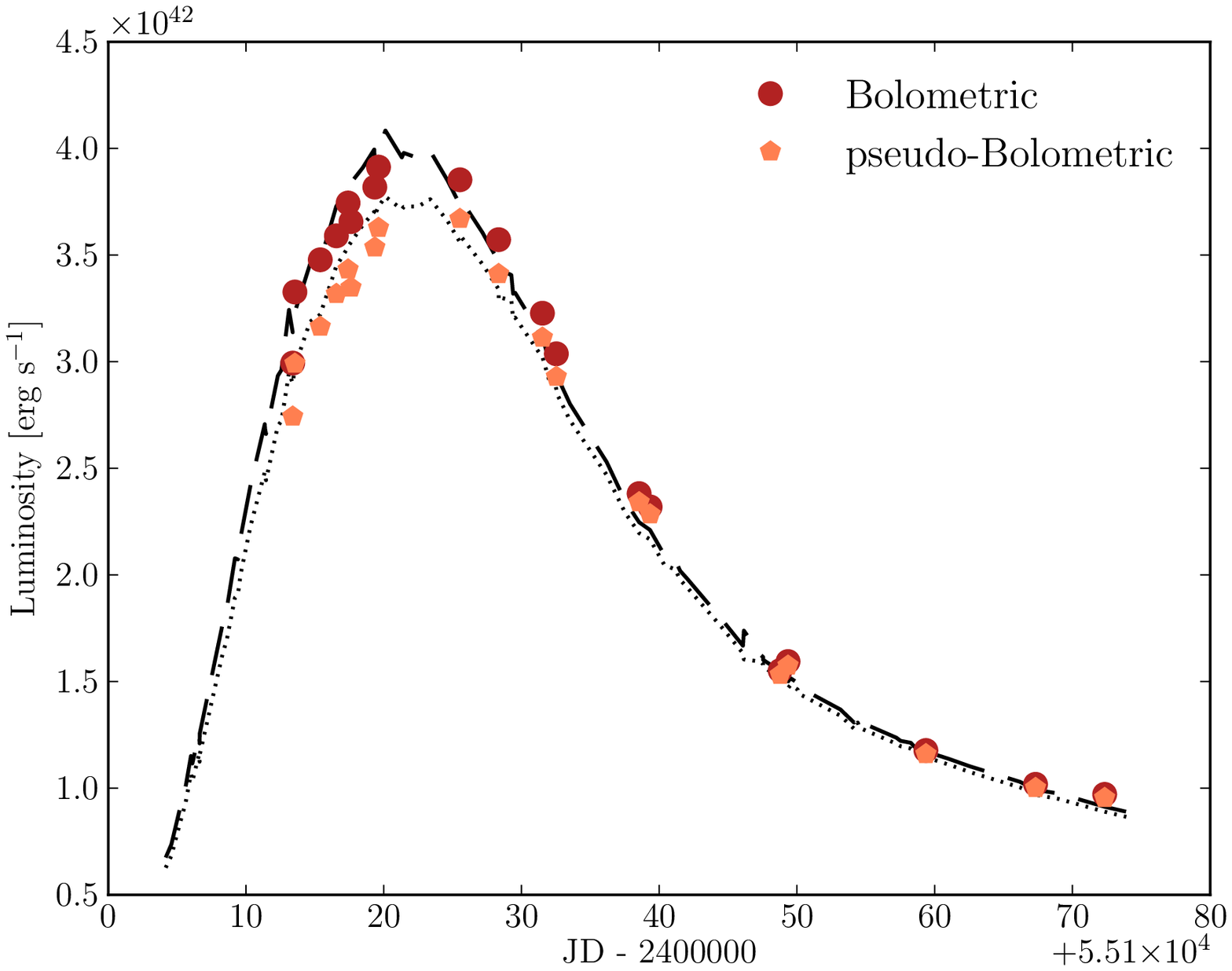}
 \caption{Comparison of the observed bolometric and pseudo-bolometric light curves found by SED integration (dashed and dotted lines respectively) with the bolometric and pseudo-bolometric light curves constructed using the BC and \pBC{} fits to $g-r$ (circles and pentagons respectively) for SN2009jf.}
 \label{fig:09jf_bolo}
\end{figure}

A comparison of the two methods' results of producing the pseudo-bolometric light curve is shown in Fig.\nobreakspace \ref {fig:09jf_bolo}. As is clear, with only $g$ and $r$ filter observations and the method presented here, an excellent estimation of the pseudo-bolometric luminosity can be obtained, even including uncertainties of extracting Sloan magnitudes from linear interpolation of a Johnson-Cousins SED (Appendix\nobreakspace \ref {sect:extractsloan}). Plotting the bolometric and pseudo-bolometric light curves highlights the contribution of the UV shown in Fig.\nobreakspace \ref {fig:fluxvstime}, with an appreciable contribution diminishing to negligible values soon after peak -- where the pseudo- and bolometric points overlap.

\label{lastpage}

\end{document}